\documentclass[prb,twocolumn,longbibliography,amsmath,superscriptaddress,amssymb,nobibnotes]{revtex4-2}

\usepackage{graphicx}
\usepackage{dcolumn}
\usepackage{bm}
\usepackage{amssymb}
\usepackage{amsmath}
\usepackage{color}
\usepackage[dvipsnames]{xcolor}
\usepackage{placeins}
\usepackage{epstopdf}
\usepackage{multirow}
\usepackage{fontenc}
\usepackage{soul}
\usepackage[normalem]{ulem}
\usepackage[linkcolor=blue,urlcolor=blue,citecolor=blue,colorlinks=true]{hyperref}

\newcommand{\drs}{{DyRh$_2$Si$_2$}}

\begin{document}

\title{Moment canting and domain effects in antiferromagnetic \drs}
\author{Kristin Kliemt}
\email{kliemt@physik.uni-frankfurt.de}
\affiliation{
 Kristall- und Materiallabor, Physikalisches Institut, 
Goethe-Universit\"at Frankfurt, 60438 Frankfurt/M, Germany
}
\author{Michelle Ocker}
\affiliation{
 Kristall- und Materiallabor, Physikalisches Institut, 
Goethe-Universit\"at Frankfurt, 60438 Frankfurt/M, Germany
}
\author{Sarah Krebber}
\affiliation{
 Kristall- und Materiallabor, Physikalisches Institut, 
Goethe-Universit\"at Frankfurt, 60438 Frankfurt/M, Germany
}
\author{Susanne~Schulz}
\affiliation{Institut f\"ur Festk\"orper- und Materialphysik, Technische Universit\"at Dresden, D-01062 Dresden, Germany}

\author{Denis~V.~Vyalikh}
\affiliation{Donostia International Physics Center (DIPC), 20018 Donostia-San Sebastian, Spain}
\affiliation{IKERBASQUE, Basque Foundation for Science, 48013 Bilbao, Spain} 

\author{Cornelius Krellner}
\affiliation{
 Kristall- und Materiallabor, Physikalisches Institut, 
Goethe-Universit\"at Frankfurt, 60438 Frankfurt/M, Germany
}

\author{Dmitry~Yu.~Usachov}
\email{dmitry.usachov@spbu.ru}
\affiliation{St. Petersburg State University, 7/9 Universitetskaya nab., St. Petersburg, 199034, Russia}
\affiliation{Moscow Institute of Physics and Technology, Institute Lane 9, Dolgoprudny, Russia}
\affiliation{National University of Science and Technology MISIS, Moscow, 119049 Russia}

\date{\today}
\begin{abstract}
A combined experimental and theoretical study of the layered antiferromagnetic compound \drs\, in the ThCr$_2$Si$_2$-type structure is presented. The heat capacity shows two transitions upon cooling, the first one at the N{\'e}el temperature $T_{\rm N}=55\,\rm K$ and a second one at $T_{\rm N2}=12\,\rm K$. Using magnetization measurements, we study the canting process of the Dy moments upon changing the temperature and can assign $T_{\rm N2}$ to the onset of the canting of the magnetic moments towards the $[100]$  direction away from the $c$ axis. Furthermore, we found that the field dependence of the magnetization is highly anisotropic and shows a two-step process for $H\parallel 001$. We used a mean-field model to determine the crystalline electric field as well as the exchange interaction parameters. Our magnetization data together with the calculations reveal a moment orientation close to the $[101]$ direction in the tetragonal structure at low temperatures and fields. 
Applying photoemission electron microscopy, we explore the (001) surface of the cleaved DyRh$_2$Si$_2$ single crystal and visualize Si- and Dy-terminated surfaces. Our results indicate that the Si-Rh-Si surface protects the deeper lying magnetically active Dy layers and is thus attractive for investigation of magnetic domains and their properties in the large family of LnT$_2$Si$_2$ materials.
\end{abstract}

\maketitle
\section{Introduction}
Ternary lanthanide-silicide compounds LnT$_2$Si$_2$ (Ln = lanthanide, T = transition metal) in the ThCr$_2$Si$_2$-type structure exhibit a large variety of physical ground states. Among them are Kondo systems \cite{Trovarelli2000, Movshovich1996, Gupta1983}, superconductors \cite{Trovarelli2000, Movshovich1996}, systems showing valence instabilities \cite{Kliemt2019a},  systems that host exotic skyrmion phases \cite{Khanh2020} and many compounds that allow for studies of  local-moment magnetism \cite{Slaski1983, Shigeoka2011, Kliemt2017, Kliemt2018}.
One member of this family, DyRh$_2$Si$_2$, came into the focus of interest again due to its recently investigated surface spin orientation \cite{Usachov2022} as well as its ultrafast magnetization dynamics \cite{Windsor2022}. 
According to magnetization and neutron diffraction investigations of polycrystalline samples, DyRh$_2$Si$_2$ ($I4/mmm$) shows antiferromagnetic (AFM) order below T$_{\rm N}=55\,\rm K$. In the structure, the Dy moments form ferromagnetic (FM) layers with Dy $4f$ moments pointing along the $c$ direction which are stacked antiferromagnetically along the c direction. Below T$_{\rm N2}=(15\pm 3)\,\rm K$, a canting away from the $c$ axis occurs amounting to a canting angle of about $(19\pm 5)^\circ$ with the $c$ axis at T$=4.2\,\rm K$ \cite{Felner1983, Melamud1984}. 
A similar case in the LnRh$_2$Si$_2$ family of this so-called "component-separated magnetic order" was found and studied in detail in the related compound HoRh$_2$Si$_2$ \cite{Shigeoka2011}. From M\"ossbauer studies on DyRh$_2$Si$_2$, the transition at T$_{\rm N2}$ was proposed to originate from the weak ordering of the Rh sublattice \cite{Felner1983} which, however, is not supported by the neutron diffraction data \cite{Melamud1984}. It has to be noted that for the related compound GdRh$_2$Si$_2$, a magnetic contribution of Rh was ruled out by an XMCD study \cite{Kliemt2017} explicitly. 
In DyRh$_2$Si$_2$, both magnetic transitions, at T$_{\rm N}$ and T$_{\rm N2}$, contribute strongly to the specific heat. The heat capacity and magnetization data were compared with the simulated data including crystalline electric field (CEF) effects in an earlier study by Takano {\it et al.} \cite{Takano1992}. 

The experimental results concerning the temperature-dependent bulk properties that are published until now \cite{Felner1983, Melamud1984} were given with large error bars only. In order to obtain further information about the magnetic ground state of the material, we re-investigate the anisotropic behavior of DyRh$_2$Si$_2$ in single crystalline form by means of detailed magnetization, electrical transport, and heat capacity measurements. Using a mean-field model, we determined the exchange interaction parameters and the CEF parameters. We provide the $\mu_0H-T$ phase diagrams for the magnetization process and study the material in low magnetic fields in order to identify signatures of the reorientation of antiferromagnetic domains in the magnetization data which allows for determining  the direction of the canting of the moments in the structure. 
We examined this material with x-ray photoemission electron microscopy (XPEEM) investigating the (001) surface of the cleaved crystal and visualizing both the Dy and Si terminations. Our results lay the foundation for future XPEEM studies, particularly regarding the characterization of the magnetic domain structure.


\section{Experiment}
Single crystals of DyRh$_2$Si$_2$ were grown from indium flux using a modified Bridgman method as described in \cite{Kliemt2019}.
High purity starting materials Dy (99.9\%, ChemPur), Rh (99.9\%, Heraeus), 
Si (99.9999\%, Wacker) 
and In (99.9995\%, Schuckard) were weighed 
in a graphite inner crucible and sealed in a niobium crucible 
under an argon atmosphere (99.999\%) subsequently. 
Dy, Rh, Si, and In were used in the ratio of Dy : Rh : Si : In = 1 : 2 : 2 : 49. The crystal growth was performed in a movable vertical furnace (GERO HTRV 70-250/18) using a maximum temperature of T$_{max}=1550^{\circ}$C, a slow cooling period with a rate of $1-4\,\rm K/h$ down to $1000^{\circ}$C followed by fast cooling to room temperature with $100\,\rm K/h$. The crystals were separated from the flux by etching in hydrochloric acid. The here described procedure yielded platelet-shaped crystals with typical dimensions of $2\,\rm mm \times 3\,\rm mm$ and a thickness of $50-100\,\mu m$.

Powder X-ray diffraction (PXRD) was performed using a Bruker D8 diffractometer (Bragg-Brentano geometry) with $\lambda = 1.5406$\,\AA\, CuK$_{\alpha}$ radiation.
PXRD data of crushed single crystals were recorded and confirmed the $I4/mmm$ tetragonal structure with the lattice parameters $a=4.0281$\,\AA\, and $c=9.9251$\,\AA\, which are in good agreement with literature \cite{Felner1984}. The chemical composition of the single crystals was checked by energy-dispersive x-ray spectroscopy (EDX). 
 The orientation of the platelet-shaped single crystals was determined using a Laue device with white x-rays from a tungsten anode. The comparison of the simulation with the recorded Laue patterns yielded that the largest face of the crystals is perpendicular to the $[001]$-direction and the large edges are $[110]$ planes.
Magnetization, heat capacity, and four-point resistivity measurements were performed down to $1.8\,\rm K$ using the commercial measurement options of a Quantum Design Physical Property Measurement System (PPMS).
XPEEM measurements have been conducted at the SPEEM end-station at the UE49-PGM beamline of the BESSY II synchrotron radiation facility operated by the Helmholtz-Zentrum Berlin (HZB) \cite{SPEEM}. A DyRh$_2$Si$_2$ single crystal with a size of about $2\times 2$ mm has been cleaved at room temperature in the preparation chamber at a pressure of about $1\cdot 10^{-10}$ mbar. After transfer into the measurement chamber, the cleaved sample was cooled to 30~K. The sample was oriented with the normal of the (001) surface to coincide with the optical axis of the microscope. In the XPEEM measurement, a $(20\times 20)\,\mu$m$^2$ region of the (001) surface has been studied using a photon energy of 170~eV and linear horizontal polarization. The measurements were performed in grazing incidence with an angle of 84$^\circ$ between the incoming light and the surface normal. XPEEM images have been recorded for kinetic energies of the photoexcited electrons ranging from 150~eV to 170~eV in steps of 0.1~eV. 


\section{Results and Discussion}

\subsection{Mean-field model for DyRh$_2$Si$_2$}

\begin{figure}[t]
\includegraphics[width=0.48\textwidth]{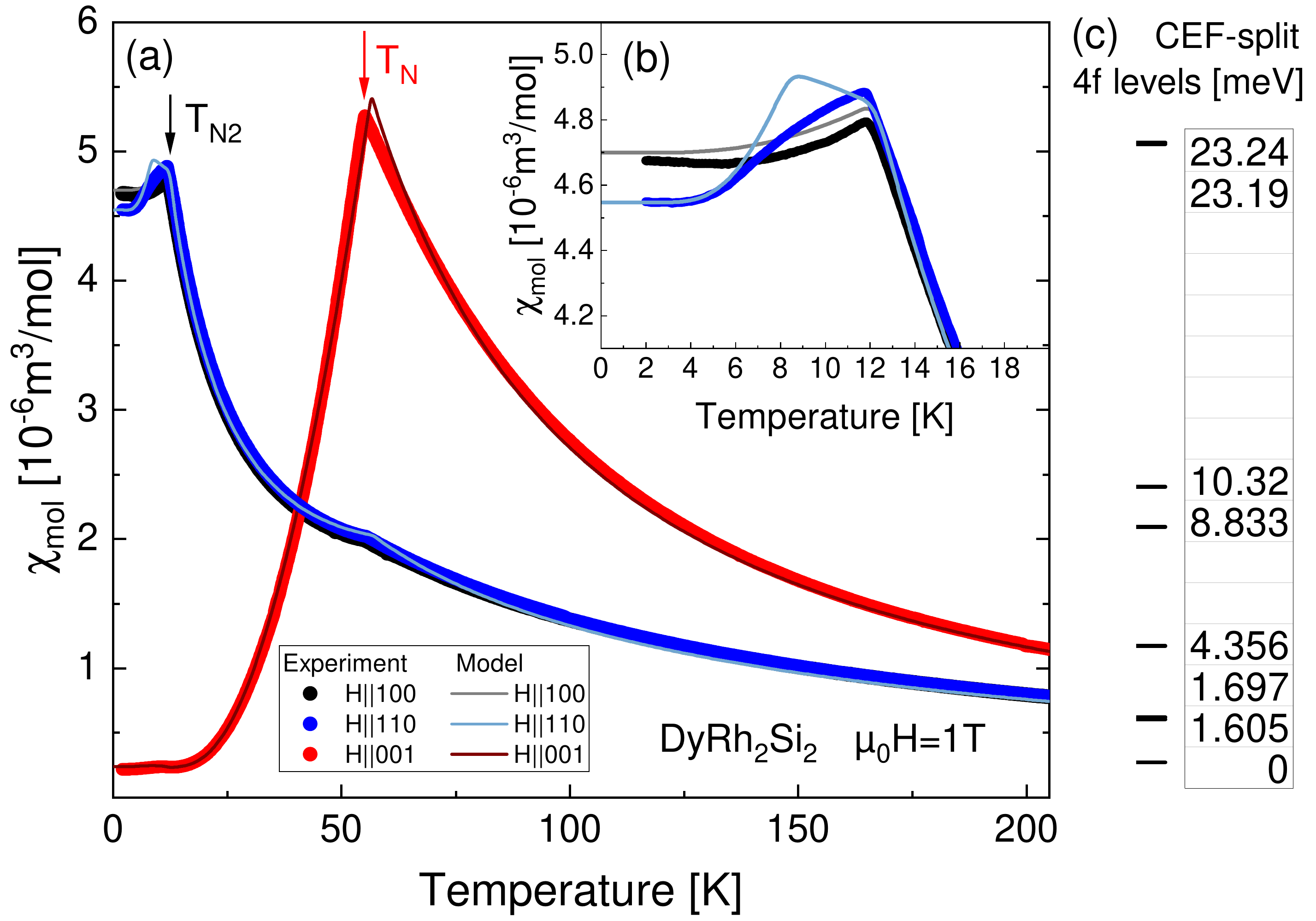}
\caption{Magnetic properties of DyRh$_2$Si$_2$. (a) Magnetic susceptibility at 1~T for three different field directions. An enlarged view of the low-temperature data is shown in the inset (b) for the in-plane fields. Solid lines result from the CEF model. (c) CEF-split 4f levels with each level being doubly degenerate. The ground state in the PM phase is an almost pure (99$\%$) $M_J = 13/2$ state.} 
\label{fig:2}
\end{figure}

\begin{figure}
\centering
\includegraphics[width=0.5\textwidth]{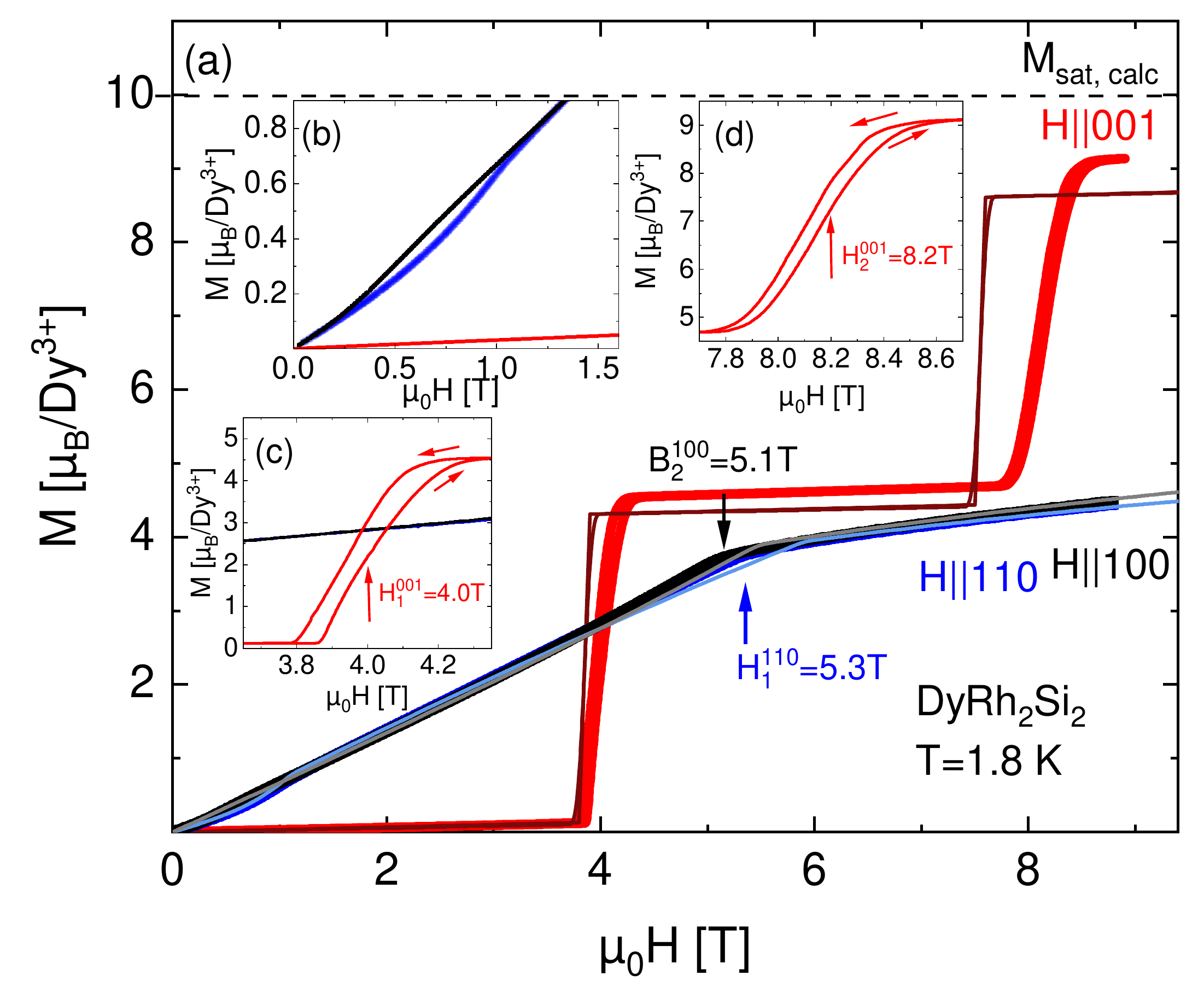}
	\caption[]{(a) Field-dependence of the magnetization per Dy$^{3+}$ at $T=1.8\,\rm K$ measured with $H\parallel 100$ (black), $H\parallel 110$ (blue) and $H\parallel 001$ (red). The simulated data are shown by grey, light blue, and brown lines. (b) For $H\parallel 100$ and  $H\parallel 110$, a reorientation of magnetic domains and a metamagnetic transition occur at small fields. Insets (c) and (d) for $H\parallel 001$, $M(H)$ shows a small hysteresis around the critical fields $\mu_0H_1^{001}$ and $\mu_0H_2^{001}$.}
\label{MvH1p8K}
\end{figure}


To simulate the magnetic properties, we used a mean-field linear-chain model with either two or four Dy sublattices. The latter was used only to model the $M(H)$ dependencies along the $[001]$ direction. The respective Hamiltonian has the form
\begin{multline}
H = \sum_{i=1}^N (H_{CEF,i} - \vec{\mu}_i\cdot\vec{B}) - \frac{3k_B}{J(J+1)} \sum_{i,j=1}^N  \\
\left[\frac{}{}T^a_{ij}J_{x,i}\langle J_{x,j}\rangle + T^a_{ij}J_{y,i}\langle J_{y,j}\rangle + T^c_{ij}J_{z,i}\langle J_{z,j}\rangle \right.\\
\left.-\frac{1}{2}\left(T^a_{ij}\langle J_{x,j}\rangle^2 + T^a_{ij}\langle J_{y,j}\rangle^2 + T^c_{ij}\langle J_{z,j}\rangle^2\right) \right],
\label{eq:Ham}
\end{multline}
where $N$ is the number of Dy sublattices, $\vec{\mu}_i$ is the magnetic moment operator, $\vec{B}$ is the external magnetic field induction, $T^a_{ij}$ and $T^c_{ij}$ are the exchange interaction parameters for the in-plane and out-of-plane directions, respectively. 
The CEF part of the Hamiltonian is
\begin{multline}
H_{CEF} = B_0^2 C_0^{(2)} + B_0^4 C_0^{(4)} + B_4^4 \left(C_4^{(4)}+C_{-4}^{(4)}\right) + \\
 B_0^6 C_0^{(6)} + B_4^6 \left(C_4^{(6)}+C_{-4}^{(6)}\right),
\label{eq:Hcef}
\end{multline}
where $B_q^k$ are CEF parameters and $C_q^{(k)}$ are spherical tensor operators. We used LSJ coupling scheme and the basis was limited to the $|M_J\rangle$ states of the ground $|LSJ\rangle$ state of the $4f^9$ subshell.

The Hamiltonian \ref{eq:Ham} was diagonalized self-consistently using an iterative scheme. We started from several intuitively selected magnetic structures and arrived at a few magnetic configurations from which the one with minimal energy was chosen. The exchange interaction parameters $T^{a(c)}_{ij}$ were fitted together with the CEF parameters $B_q^k$ to obtain best agreement between the theoretical and experimental dependencies $\chi(T)$ at 1~T (Fig.~\ref{fig:2}(a)) and $M(H)$ at 1.8~K (Fig.~\ref{MvH1p8K}) for the $[100]$, $[110]$ and $[001]$ directions. Due to symmetry, we have $T^{a(c)}_{12}=T^{a(c)}_{14}$. Additionally, the parameter $T^a_{13}$ was kept equal to $T^c_{13}$ since it cannot be determined from our data due to its linear correlation with $T^a_{11}$.

\begin{table}[ht]
\begin{center}
\begin{tabular}{ c|c|c|c|c|c|c } 
$B_0^2$ & $B_0^4$ & $B_4^4$ & $B_0^6$ & $B_4^6$ & Plane & Source\\
\hline 
$36.86$ & $-39.45$ & $44.48$ & $0.837$ & $11.98$ & $(100)$ & this work\\
$31.76$ & $-37.85$ &  &  &  & undefined & Ref.~\cite{Tomala1989} \\
$36.65$ & $-50.07$ & $6.959$ &  &  & $(100)$ & Ref.~\cite{Takano1992}\\
\end{tabular}
\end{center}
\caption{CEF parameters $B^k_q$ in meV and corresponding crystallographic plane, in which the moments deviate from the $[001]$ direction.}
\label{tab:1}
\end{table}

\begin{table}[ht]
\begin{center}
\begin{tabular}{c|c|c|c|c} 
$T^a_{11}$ & $T^a_{12}=T^a_{14}$ & $T^a_{13}=T^c_{13}$ & $T^c_{11}$ & $T^c_{12}=T^c_{14}$\\
\hline 
 $24.907$~K & $-9.194$~K & $-5.5$~K & $23.049$~K & $-8.245$~K  \\
\end{tabular}
\end{center}
\caption{Optimized exchange interaction parameters of the mean-field model.}
\label{tab:2}
\end{table}

The obtained CEF parameters are given in Tab.~\ref{tab:1}. It should be noted that in the previous studies \cite{Tomala1989, Takano1992} the parameters $B_0^6$ and $B_4^6$ were not estimated. The $B_0^2$ and $B_4^0$ parameters show a reasonable agreement with our values. The exchange interaction parameters given in Tab.~\ref{tab:2} were found to be slightly different for the in-plane ($a$) and out-of-plane ($c$) directions. Without this anisotropy, the simulated magnetic properties would show significantly worse agreement with the experimental data.
It should be noted that although our model provides a good fit for the susceptibility in Fig.~\ref{fig:2}, one can see some discrepancies close to $T_{\rm N}$ in the $[001]$ direction and close to $T_{\rm N2}$ in the $[110]$ direction. This is probably related to the neglect of spin fluctuations.

\subsection{Magnetization}
The magnetic properties of DyRh$_2$Si$_2$ arise from the local moment magnetism of the Dy$^{3+}$ ion with a spin momentum of $S=5/2$ and an orbital angular momentum $L=5$ resulting in a total angular momentum of $J=15/2$.  
 Polycrystalline samples of the system showed AFM order below T$_N=55\,\rm K$ with moments along $[001]$ \cite{Felner1983, Melamud1984}. A second transition was found at $\approx 15\,\rm K$ which was assigned to the onset of a canting process of the magnetic moments away from the $[001]$ direction. Our single crystals allow for a detailed study of the anisotropic magnetization of DyRh$_2$Si$_2$. We determined the temperature and field-dependent magnetic susceptibility with field along the three crystallographic main symmetry directions $[001]$, $[100]$ and $[110]$ in order to obtain further information about the magnetic ground state of the material, especially about the orientation of the magnetic moments in the bulk.

\subsubsection{Temperature dependence of the susceptibility}
\begin{figure}
\centering
\includegraphics[width=0.45\textwidth]{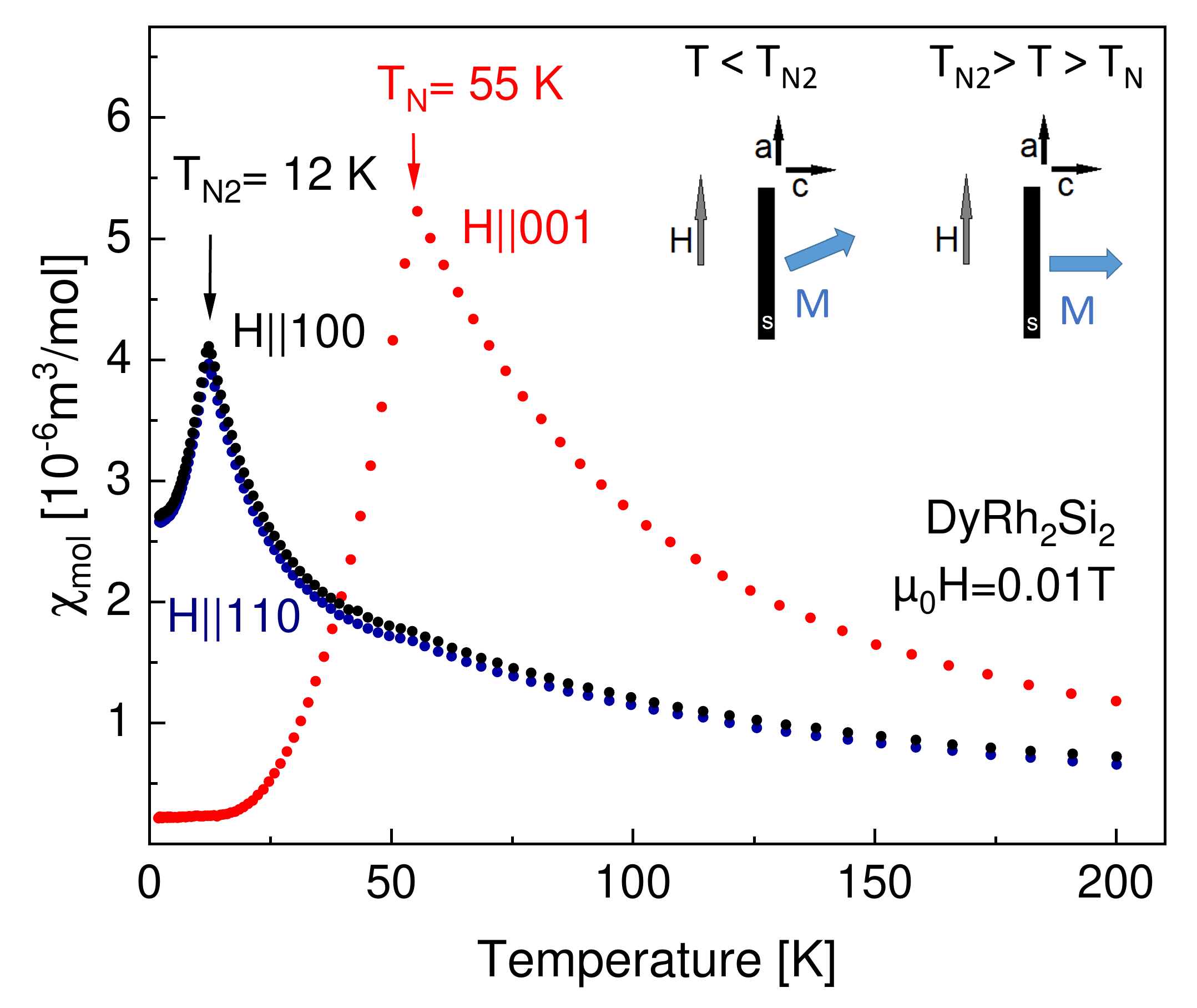} 
	\caption[]{Temperature-dependent susceptibility for the three main symmetry directions with $\mu_0H=0.01\rm\, T$. In the inset, a schematic drawing of the experimental setup is shown for $H\perp 001$ of the sample $s$.}
\label{MvT_0p01T}
\end{figure}

In Fig.~\ref{MvT_0p01T}, the comparison of the anisotropic temperature dependent susceptibility, $\chi_{\rm mol}(T)$, measured at $0.01\,\rm T$ is presented for $H\parallel100, H\parallel110$ and $H\parallel001$.
$\chi_{\rm mol}(T)$ for $H\parallel 001$, shows that the system enters the antiferromagnetic phase below $T_{\rm N}= 55\,\rm K$. 
In contrast, when the field is applied along the $[100]$ or $[110]$ directions (inset of Fig.~\ref{MvT_0p01T}), $\chi_{\rm mol}(T)$ shows only a small kink at $T_{\rm N}$ but a 
pronounced anomaly followed by a strong decrease below $T_{\rm N2}= 12\,\rm K$. In our setup, we measure the projection of the Dy moment $M$ onto the direction of the applied field $H$, ${\bf M\cdot H}/H$ and the almost complete absence of a kink at $T_{\rm N}$ for $H\perp 001$ is consistent with moment alignment along $c$ at the Neel temperature for which ${\bf M\cdot H}/H = 0$.  Below $T_{\rm N2}$, the moments tilt away from the $c$ axis leading to a non-zero projection for $H\perp 001$ and a strong change of the susceptibility.
For different higher fields applied along the three main symmetry directions, $\chi_{\rm mol}(T)$ is shown in the Supporting Information 
 Figs.~\ref{MvT001}(a),\ref{MvT100}(a),\ref{MvT110}(a).

\begin{figure}
\centering
\includegraphics[width=0.45\textwidth]{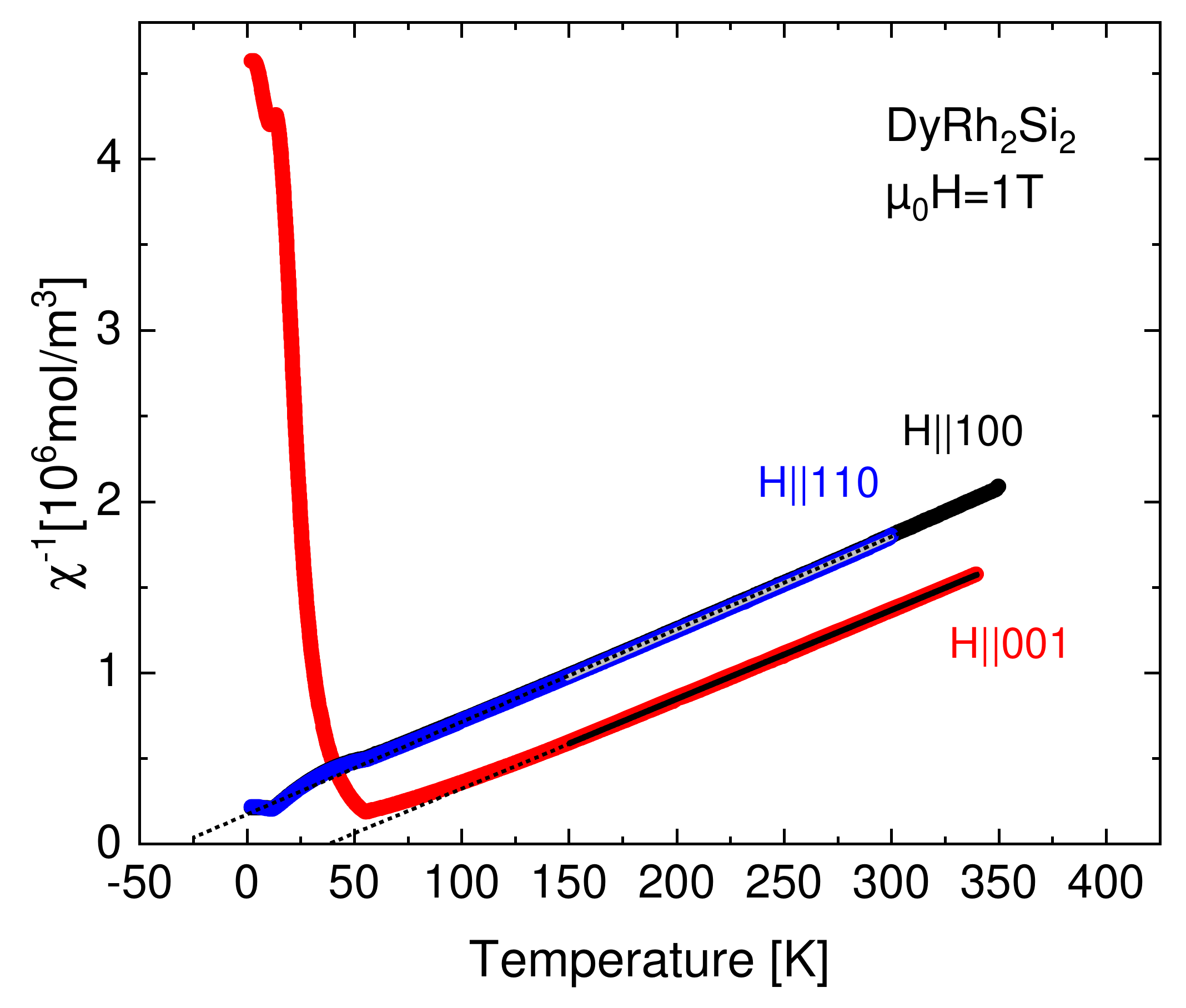}
	\caption[]{Inverse susceptibility measured at $\mu_0H=1\,\rm T$. Solid gray and black lines indicate the region of the Curie-Weiss fit above $150\,\rm K$.}
\label{Chi-1vT}
\end{figure}
 Following the common practice, we determined the Weiss temperature $\Theta_{\rm W}$ and the effective magnetic moment $\mu_{\rm eff}$  according to
\begin{eqnarray}
\chi^{-1}_{\rm mol}(T)=-\frac{\Theta_W}{C_{\rm mol}}+\frac{1}{C_{\rm mol}}T
\end{eqnarray} 
and 
$\mu_{\rm eff}=\sqrt{{3 k_{\rm B} C_{\rm mol}}/{\mu_0 N_A}}$
with the Boltzmann constant $k_{\rm B}$, the molar Curie constant $C_{\rm mol}$, the vacuum permeability $\mu_0$ and the Avogadro number $N_A$ from the inverse susceptibility, Fig.~\ref{Chi-1vT}, measured with $\mu_0H=1\rm\,T$. 
We obtained negative Weiss temperatures  $\Theta^{100}_{\rm W}=-29.4\,\rm K$, $\Theta^{110}_{\rm W}=-30.2\,\rm K$ for in-plane fields and a positive Weiss temperature, $\Theta^{001}_{\rm W}=36.5\,\rm K$, for $H\parallel 001$.
The main reason for the different signs of the Weiss temperatures is related to the CEF that leads to the strong anisotropy of the susceptibility.
The effective magnetic moment $\mu^{100,110}_{\rm eff}=10.8\,\mu_{\rm B}$ and $\mu^{001}_{\rm eff}=11.1\,\mu_{\rm B}$ is slightly anisotropic and agrees well with the calculated value 
$\mu_{\rm eff}^{\rm calc}= g_J\sqrt{J(J+1)}\,\mu_{\rm B}= 10.65\,\mu_{\rm B}$ which is the highest among all Ln$^{3+}$ ions.




\subsubsection{Field-dependence of the magnetization}

\begin{figure}
\centering
\includegraphics[width=0.5\textwidth]{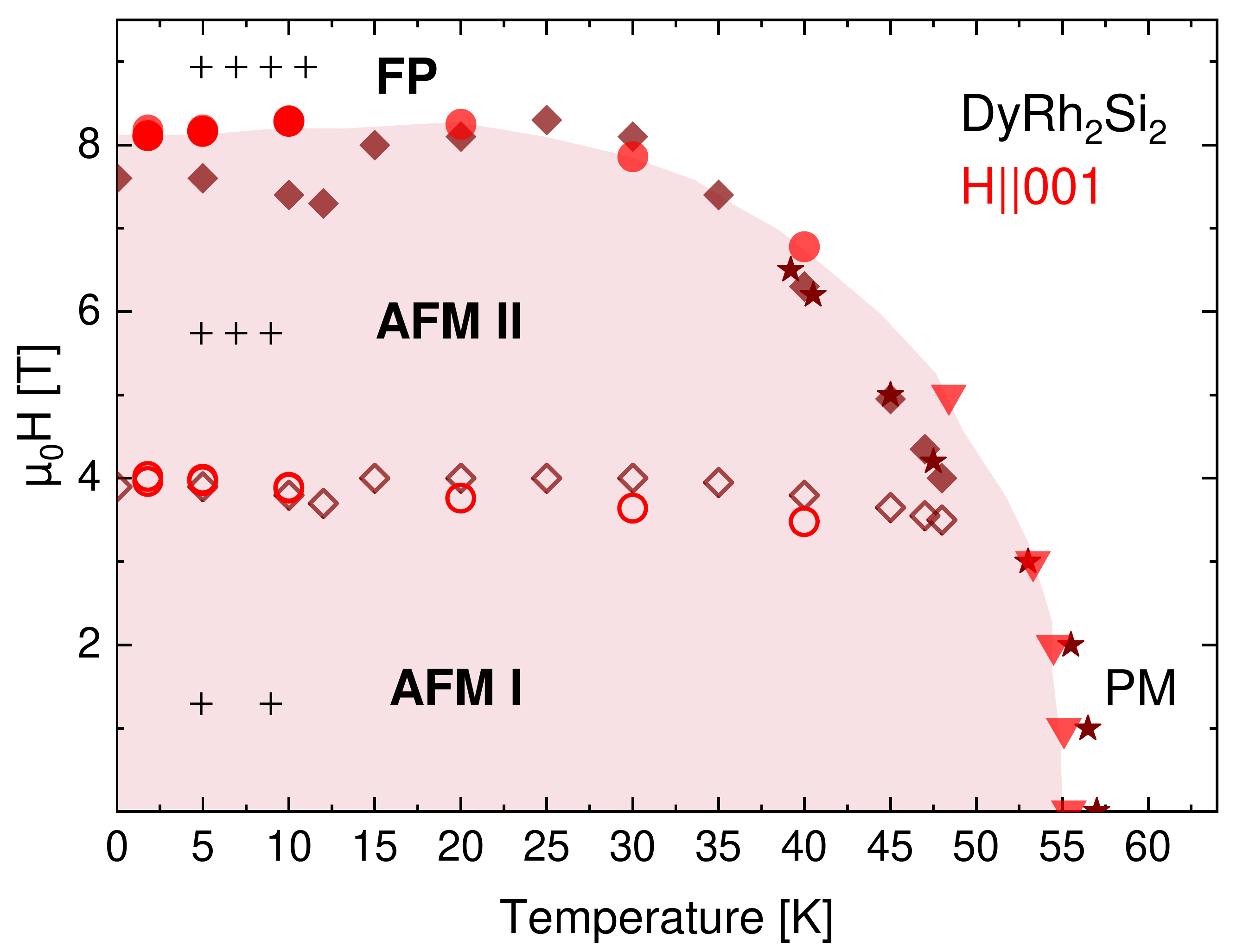}
	\caption[]{$\mu_0H-T$ phase diagram for $H\parallel 001$. In DyRh$_2$Si$_2$, upon increasing the field, the magnetic configuration changes in two steps from $+-+-$ (AFM I) to $+++-$ (AFM II) before reaching the $++++$ field polarized (FP) state. Upon increasing the temperature, the paramagnetic (PM) state is reached. Red circles mark transitions extracted from $M(H)$ data and triangles were extracted from $\chi_{\rm mol}(T)$. Brown diamonds and stars show the result of the simulation of $M(H)$ and $\chi_{\rm mol}(T)$.}
\label{B-T_diagram001}
\end{figure}

\begin{figure}
\centering
\includegraphics[width=0.5\textwidth]{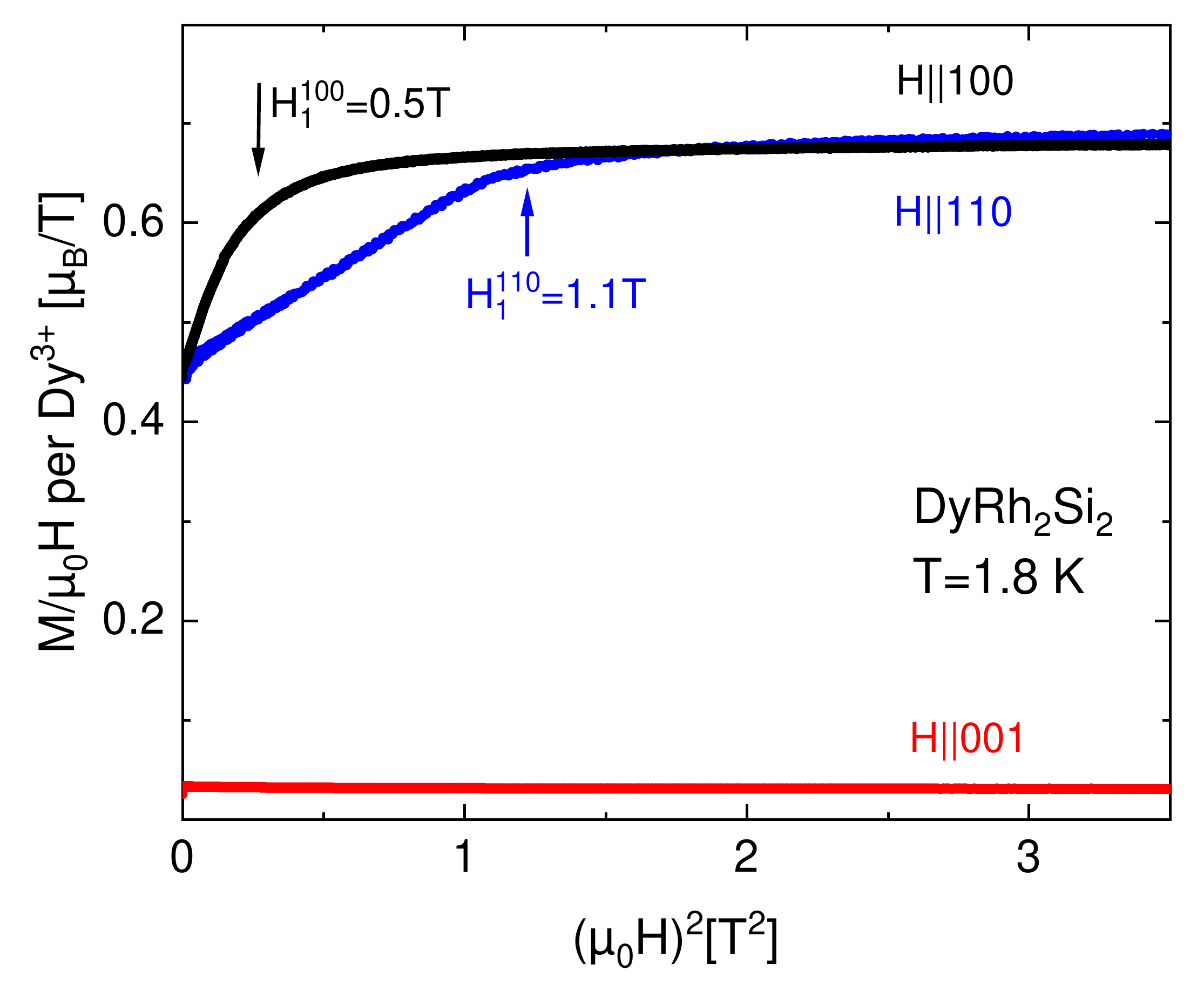}
	\caption[]{Comparison of the low-field behavior of ${\bf M\cdot H}/H^2$ vs. $(\mu_0H)^2$ at $T=1.8\,\rm K$ for increasing $H$ along the three  main symmetry directions.}
\label{MvH_Lowfield100}
\end{figure}
The field-dependence of the magnetization at $T=1.8\,\rm K$ shows anisotropic behavior and is presented in Fig.~\ref{MvH1p8K} (a). For $H\parallel 001$ (red curve), we observe a step-like behavior of the magnetization due to  changes of the magnetic configuration at $\mu_0H_1^{001}= 4.0\,\rm T$ and at $\mu_0H_2^{001}= 8.2\,\rm T$. At both transitions, a small hysteresis in the magnetization occurs Fig.~\ref{MvH1p8K} (c) and (d). With $M=9.1\,\mu_{\rm B}/\rm Dy^{3+}$ at the highest here investigated field, the magnetization is slightly lower than the calculated saturation magnetization of $M_{\rm sat}^{\rm calc}=g_JJ=10\mu_{\rm B}/\rm Dy^{3+}$ with $g=4/3$ being the Land{\'e} $g$ factor. The reason for this is the deviation of moments from the $c$ axis due to the CEF. In our model, the field of 36~T is required to align all moments along the $c$ direction at 1.8~K.
At $\mu_0H_1^{001}= 4.0\,\rm T$ the magnetic configuration
changes from $+-+-$ (AFM I) to $+++-$ (AFM II) and at $\mu_0H_2^{001}= 8.2\,\rm T$ to the field-polarized $++++$ (FP) configuration.
Our mean-field simulations indicate that the reason for such a two-step process is related to the negative sign of the $T_{13}^c$ parameter. Similarly to the case of TbRh$_2$Si$_2$~\cite{Abe_JPSJ_2002}, AFM coupling between the neighbor Dy sublattices (layers) 1 and 2 is in conflict with the AFM coupling between the sublattices 1 and 3. As a result of such frustration, the moments in the two sublattices flip at different values of $H$. In the case of positive $T_{13}^c$ there would be only one step.
Next, we extracted the reorientation fields at different temperatures from Figs.~\ref{MvT001}(a) and \ref{MvH001}(a) 
in the Supporting Information to draw the $\mu_0H-T$ phase diagram for $H\parallel 001$, as shown in Fig.~\ref{B-T_diagram001}. Our experimental diagram is qualitatively similar to that of TbRh$_2$Si$_2$~\cite{Abe_JPSJ_2002}. However, our mean-field model gives better agreement with experiment than the 1D Ising chain model used in Ref.~\cite{Abe_JPSJ_2002}.

For $H\parallel 100$ (black curve in Figs.~\ref{MvH1p8K} and \ref{MvH_Lowfield100}) and $H\parallel 110$ (blue curve), we observe changes of the slope at $\mu_0H_1^{100}=0.5\,\rm T$, $\mu_0H_2^{100}= 5.1\,\rm T$, $\mu_0H_1^{110}= 1.1\,\rm T$ and at $\mu_0H_2^{110}= 5.3\,\rm T$ as shown in Fig.~\ref{MvH1p8K} (a,b) and Fig.~\ref{MvH_Lowfield100}. For in-plane fields, the magnetization at $8\,\rm T$ is much lower than the calculated saturation value.

\subsubsection{Low-field behavior of the magnetization}
\begin{figure}
\centering
\includegraphics[width=0.5\textwidth]{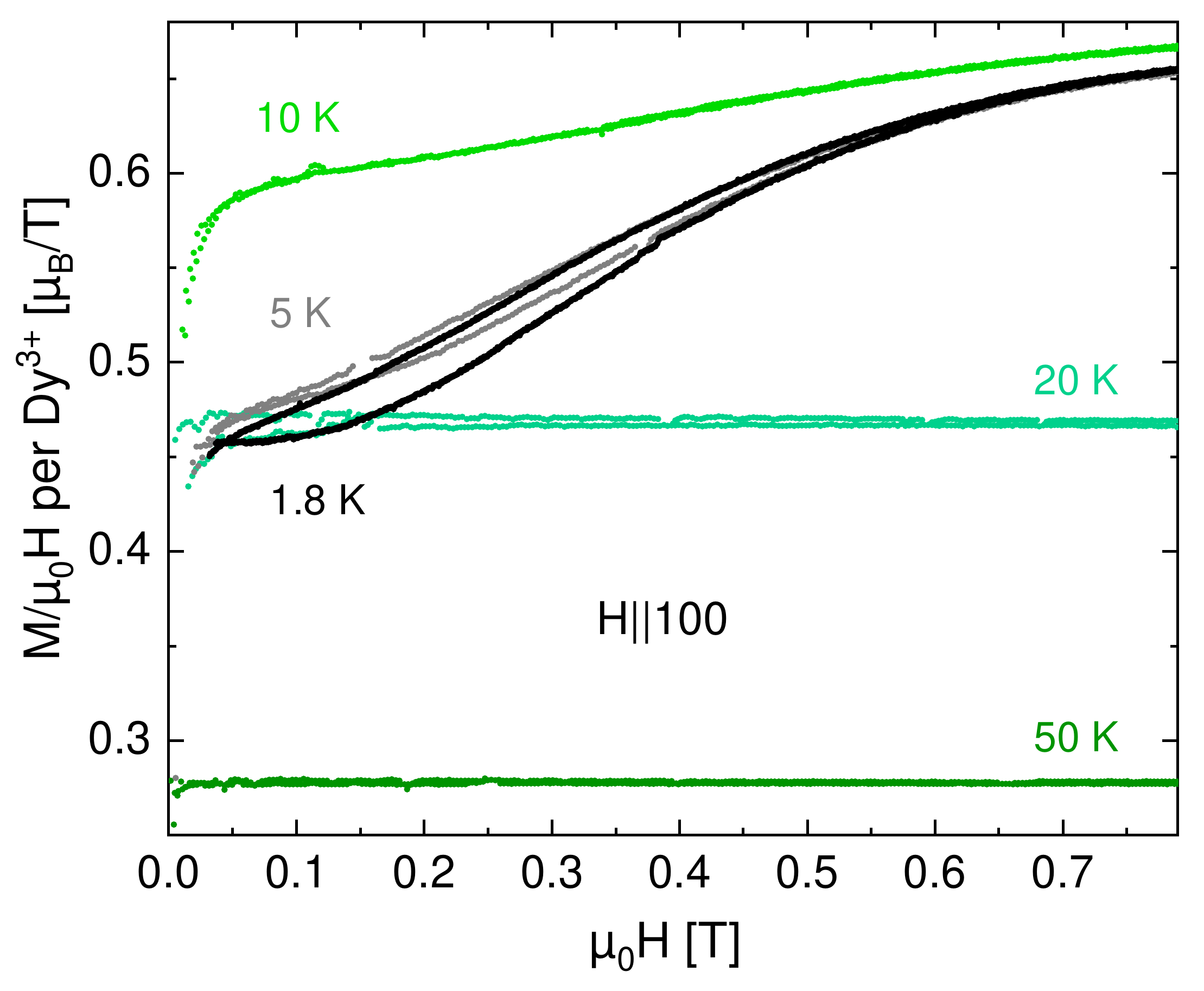}
	\caption[]{${\bf M\cdot H}/H^2$ versus $\mu_0H$ for $H\parallel 100$. Below $10\,\rm K$, a strong change of slope and a small AFM hysteresis occur which indicates the reorientation of magnetic domains.}
\label{Fig6_Dom100}
\end{figure}

By comparing the low-field behavior of the magnetization per Dy$^{3+}$ for the three main symmetry directions, presented as $M/\mu_0H$ versus $(\mu_0H)^2$ in Fig.~\ref{MvH_Lowfield100}, we find an anisotropy not only between the in and out-of-plane data but also between the in-plane data (black curve $H\parallel100$ and blue curve $H\parallel110$). 
 A closer look on the low-field data for $H\parallel100$, shown in  Fig.~\ref{Fig6_Dom100}, most clearly visible when displayed as $M/\mu_0H$ versus $\mu_0H$, indicates that below $10\,\rm K$ strong changes of $M/\mu_0H$ appear at fields below $0.5\,\rm T$. The curves recorded at $1.8\,\rm K$ and $5\,\rm K$ show a small field-induced hysteresis. We deduce that this change of the slope for $H\parallel100$ at low fields is connected to a magnetic reorientation below $\mu_0H_1^{100}=0.5\,\rm T$. A similar non-linear behavior was found in GdRh$_2$Si$_2$ ($H\parallel110$) and was assigned to the reorientation of antiferromagnetic domains \cite{Kliemt2017}. Note that our mean-field model does not describe the reorientation of domains. In contrast, for $H\parallel 110$ (blue curve), $M\propto (\mu_0H)^2$ and a metamagnetic spin-flop transition occurs below $\mu_0H_1^{110}= 1.1\,\rm T$ (Figs.~\ref{MvH_Lowfield100} and 
 ~\ref{MvH110_Spinflop}) 
 that is well-described by our model. For $H\parallel 001$ (red curve) in Fig.~\ref{MvH1p8K} (a), the susceptibility is almost zero below $\mu_0H_1^{001}$ and together with the domain-reorientation behavior observed for $H\parallel 100$ below  $\mu_0H_1^{100}=0.5\,\rm T$, this indicates a moment tilting towards the $[100]$ direction which is consistent with the result of the calculations given in Tab.~\ref{tab:1}. Our calculations show that the moments are oriented along $[001]$ between $T_{\rm N}$ and $T_{\rm N2}$ and become canted in the $(100)$ or $(010)$ plane at lower temperatures and zero external fields with the maximal canting angle of $24.3^\circ$ from the $c$ direction. The field dependence of the material was experimentally studied with $H\parallel 100$, for different higher temperatures and is compared to the simulation in  Fig.~\ref{MvH100} 
 in the Supporting Information.

\begin{figure}
\centering
\includegraphics[width=0.5\textwidth]{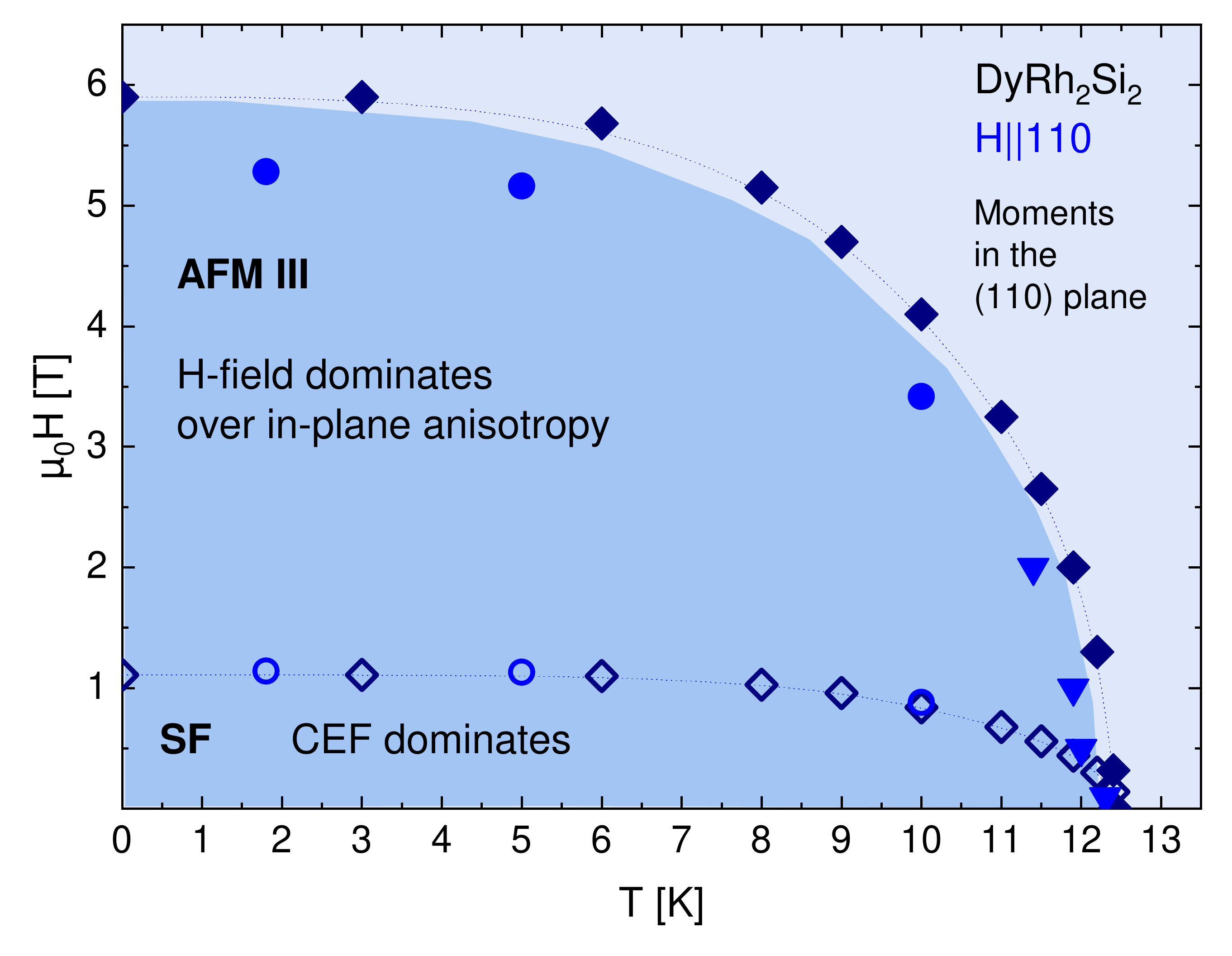}
	\caption[]{$\mu_0H-T$ phase diagram for $H\parallel 110$. Circles mark transitions extracted from $M(H)$ data and triangles were extracted from $\chi_{\rm mol}(T)$. Diamonds show the results of the simulations.}
\label{B-T_diagram110}
\end{figure}

For $H\parallel 110$, $\mu_0H<1.3\,\rm T$ and low temperatures, 
Fig.~\ref{MvH110_Spinflop}, 
$M/H$ shows a linear $(\mu_0H)^2$ dependence as the signature of a spin-flop transition \cite{Kliemt2017}. In zero field and below $T_{\rm N2}$ the moments are oriented by the CEF in the $(100)$ (or $(010)$) plane. Upon the increase of the field, they tend to be orthogonal to the field. This results in the gradual rotation of the moments towards the $(110)$ plane.
From the field- and temperature-dependent data Figs.~\ref{MvT110},\ref{MvH110_Spinflop}, and \ref{MvH110_all}(a)
(Supporting Information), we extracted the transition temperatures in the field along $[110]$ to draw the $\mu_0H-T$ phase diagram, Fig.~\ref{B-T_diagram110}, in which the region of the spin-flop (SF) phase is marked.

\subsection{Heat capacity and entropy}

\begin{figure}
\centering
\includegraphics[width=0.5\textwidth]{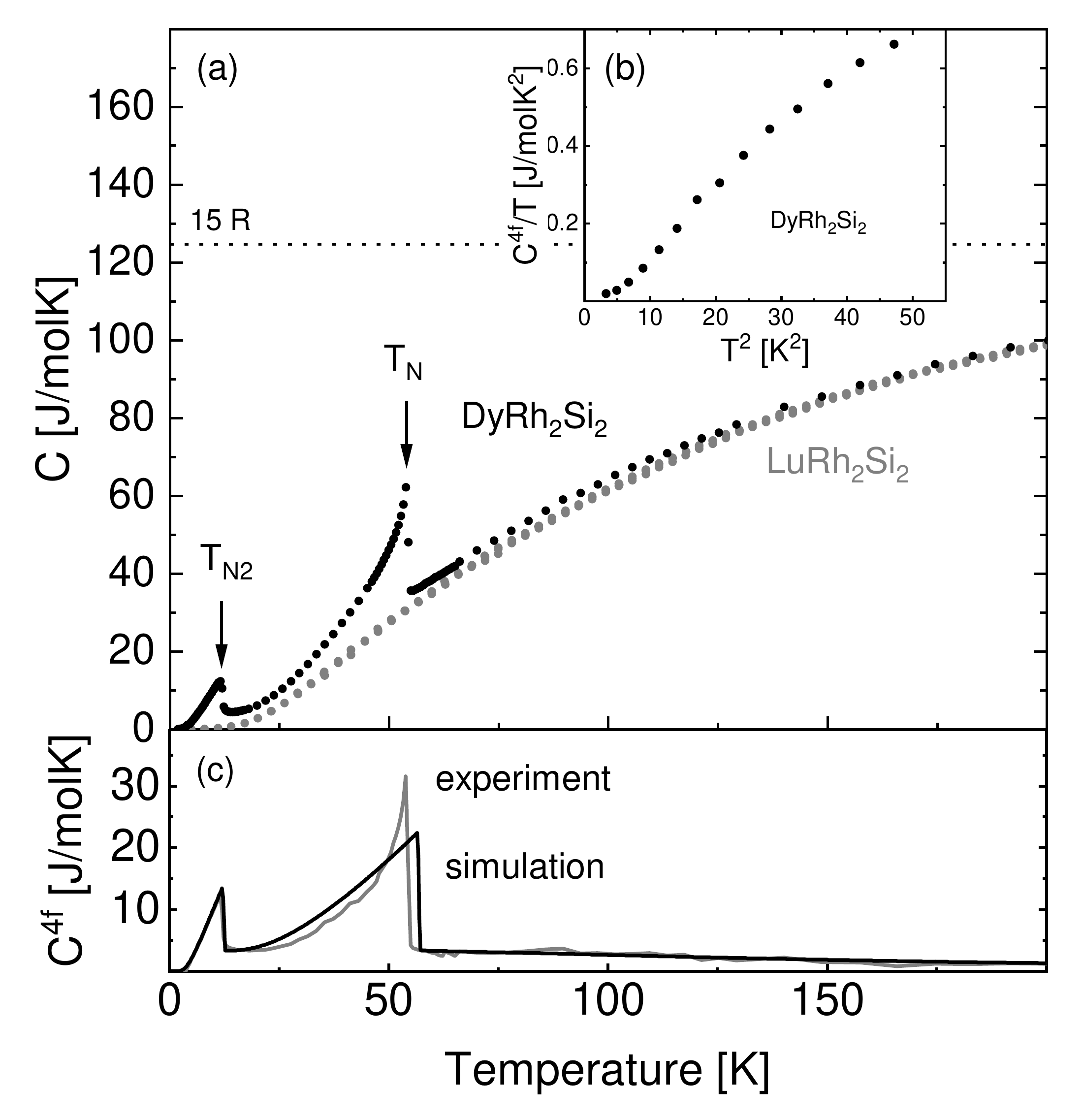}
	\caption[]{(a) Heat capacity of DyRh$_2$Si$_2$ and LuRh$_2$Si$_2$ \cite{Ferstl2007}. (b) $\rm C/T$ versus T$^2$ of DyRh$_2$Si$_2$ does not show linear behavior below $10\,\rm K$ which hinders the determination of the Sommerfeld coefficient and hints at a magnon contribution to the heat capacity below T$_{\rm N2}$. (c) The magnetic part of the heat capacity, $C^{4f}$, was determined by using  the non-magnetic reference LuRh$_2$Si$_2$ (gray symbols in (b)).}
\label{HC_DyRh2Si2}
\end{figure}

The heat capacity, Fig.~\ref{HC_DyRh2Si2}, was measured between $1.8\,\rm K$ and $200\,\rm K$. It shows a $\lambda$-type anomaly at $T_{\rm N}=55\,\rm K$ and a further transition at $T_{\rm N2}=12\,\rm K$.
The recorded data agree well with those published by Takano \textit{et al.} \cite{Takano1992}.
The phonon contribution was subtracted using 
the nonmagnetic reference data of LuRh$_2$Si$_2$ which were taken from \cite{Ferstl2007}. We note that below $5\,\rm K$ the Lu reference exhibits a larger heat capacity than DyRh$_2$Si$_2$ which leads to unphysically negative values of the magnetic specific heat and of the entropy at low temperatures. Possible reasons might be a higher electronic heat capacity or a slightly different phonon spectrum in LuRh$_2$Si$_2$ due to the mass differences of Dy and Lu.
 Due to the low-lying transition at $T_{\rm N2}$ and magnonic degrees of freedom, $C/T$ versus $T^2$ shows a non-linear behavior that hinders the determination of the Sommerfeld coefficient $\gamma$ from the low-temperature data.
\begin{figure}
\centering
\includegraphics[width=0.5\textwidth]{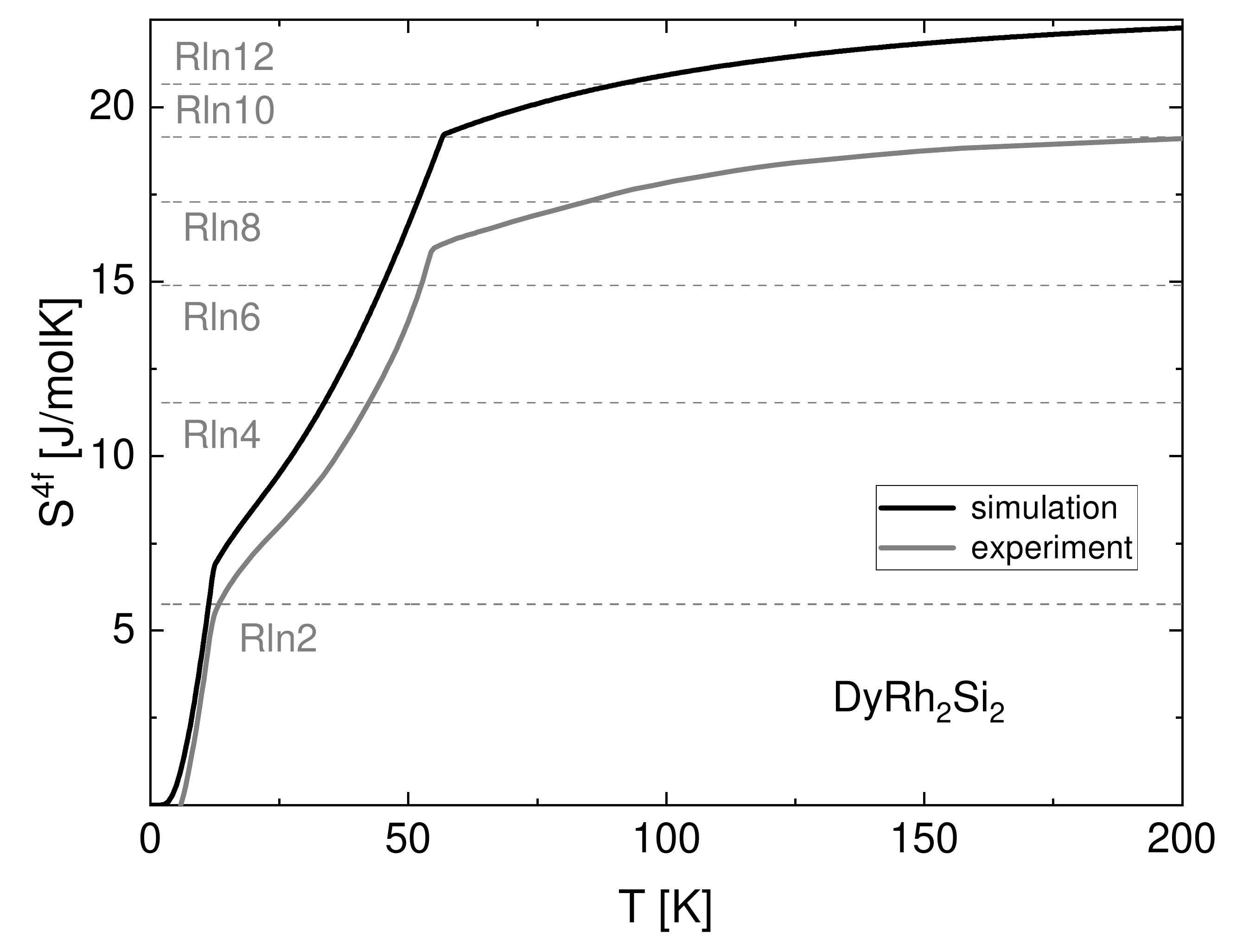}
	\caption[]{Magnetic entropy, S$^{4f}$, of DyRh$_2$Si$_2$. The experimental data were obtained from C$^{4f}$ after subtracting the data of the non-magnetic reference (gray solid line). The simulation is shown in black.}
\label{Entropy_DyRh2Si2}
\end{figure}
The magnetic entropy, Fig.~\ref{Entropy_DyRh2Si2}, of DyRh$_2$Si$_2$ reaches almost 
$S^{4f}=\rm R\ln 8$ at T$_{\rm N}$ which is much smaller than the expected entropy $S^{4f}(\rm Dy^{3+})=\rm R\ln(2J+1)=\rm R \ln 16$ for the free Dy$^{3+}$ ion, this indicates  that the overall splitting of the CEF levels is larger than T$_{\rm N}$ which is consistent with our calculated CEF scheme.
From Fig.~\ref{fig:2}(c) it can be seen, that below $T_{\rm N}=55\,\rm K\approx 4.7\,\rm meV$ the lowest four doubly degenerate CEF energy levels have a significantly greater population than the higher levels. This is in agreement with the entropy being close to $S^{4f}=\rm R\ln 8$ for the 4 CEF doublets.


\subsection{Electrical resistivity}
\begin{figure}
\centering
\includegraphics[width=0.5\textwidth]{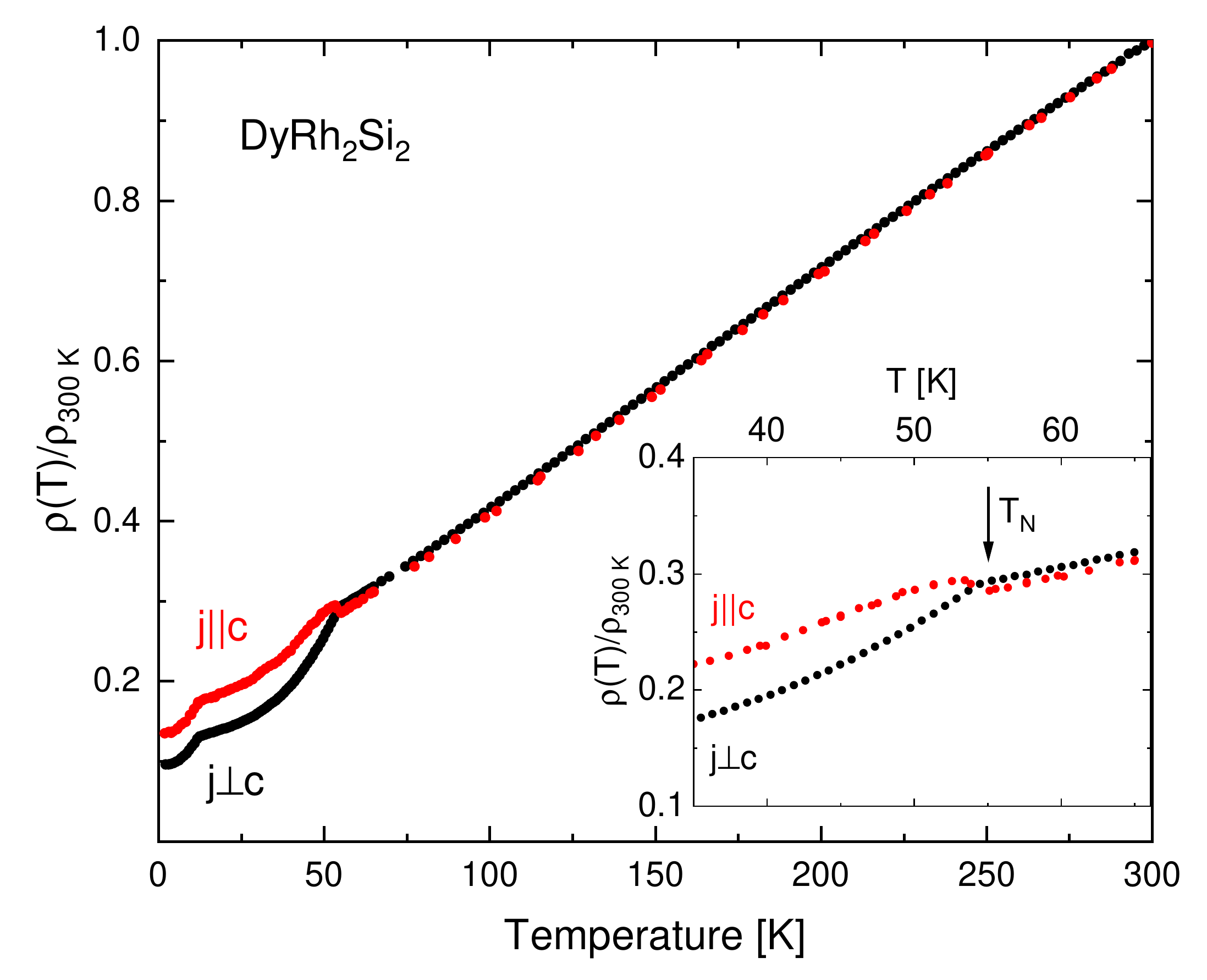}
	\caption[]{Normalized resistivity of DyRh$_2$Si$_2$}
\label{Rho_DyRh2Si2}
\end{figure}
The normalized electrical resistivity, Fig.~\ref{Rho_DyRh2Si2}, exhibits linear behavior for both current directions between $300\,\rm K$ and T$_{\rm N}$. At T$_{\rm N}$, upon entering the antiferromagnetic phase the behavior is different for both current directions $j\perp c$ and $j\parallel c$. As shown in the inset of Fig.~\ref{Rho_DyRh2Si2}, below T$_{\rm N}$ the resistivity drops for $j\perp c$ while it shows a slight increase for $j\parallel c$ indicating a decrease of the number of states near the Fermi level caused by a sudden slight enlargement of the unit cell in $c$-direction. It has to be noted that for the sister compound GdRh$_2$Si$_2$ a similar but even stronger effect on the temperature dependence of the resistivity can be observed \cite{Kliemt2015}. There, this increase of the resistivity for $j\parallel c$ occurs below $T_{\rm N}$ where the $c$ parameter of the magnetic unit cell becomes twice as large as that of the structural cell  which reduces the number of states that contribute to the conductivity. 
The residual resistivity ratio of 
RR$_{2\,\rm K}=10.4$ for $j\perp c$ is slightly larger then RR$_{2\,\rm K} = 7.4$ determined for $j\parallel c$.

\subsection{XPEEM}

\begin{figure*}
\centering
\includegraphics[width=\textwidth]{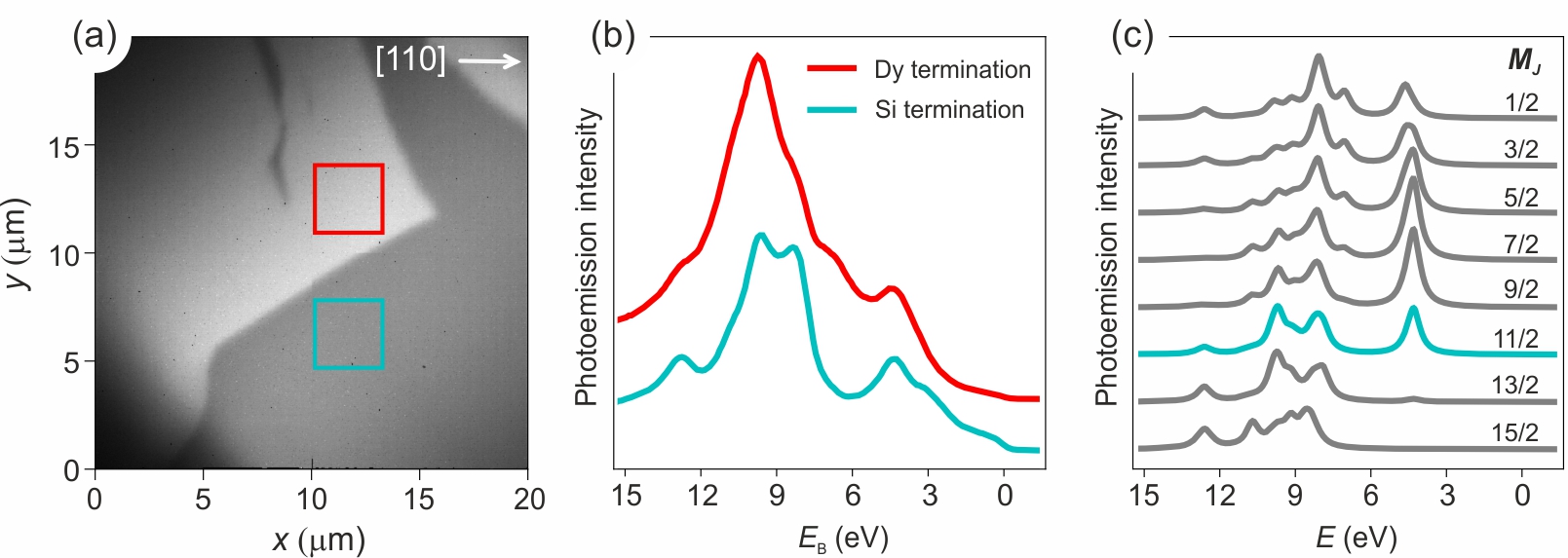}
	\caption[]{Spectromicroscopic insight into the surface terminations of DyRh$_2$Si$_2$. (a) Representative XPEEM image showing a $(20 \times 20)\,\mu$m$^2$ area of the (001) surface from a cleaved DyRh$_2$Si$_2$ single crystal at a temperature of 30~K using a photon energy of 170~eV. The image results from integration over kinetic energies of the photoelectrons ranging from 150~eV to 170~eV. Bright and dark regions reflect surface areas with Dy and Si terminations, respectively. (b) Dy~4$f$ photoemission spectra derived from the bright and dark areas in the XPEEM image that are highlighted by red- and cyan-colored squares, respectively.  (c) Calculated $M_J$-dependent Dy~4$f$ spectra for the geometry of the XPEEM measurements. The calculated spectrum that shows the best agreement with the experimental one taken from the Si surface is highlighted in cyan.}
\label{PEEM_DyRh2Si2}
\end{figure*}

We now discuss the spectromicroscopic investigation of the cleaved (001) surface of AFM-ordered DyRh$_2$Si$_2$. The results are summarized in Fig.~\ref{PEEM_DyRh2Si2}. 
The ultimate XPEEM image shown in Fig.~\ref{PEEM_DyRh2Si2}(a) results from integration over the given kinetic energies.
As we can see from Fig.~\ref{PEEM_DyRh2Si2}(a), the XPEEM image taken with a photon energy of 170~eV for kinetic energies ranging from 150~eV to 170~eV, reveals large and homogeneous, bright- and dark-shaded areas. At the given photon energy, the cross-section for the photoexcitation of valence electrons is considerably smaller than for the excitation of electrons from the Dy~4$f$ shell. From ARPES studies of DyRh$_2$Si$_2$ and related LnRh$_2$Si$_2$ compounds, it is known, that cleaving of the sample results in two different terminations of the (001) surface: Dy or Si termination. Because of the high surface sensitivity at the given photon energy due to the small escape depth of the photoelectrons, which decays exponentially with increasing distance from the surface, we may assume that the more intense, bright areas in the XPEEM image reflect Dy-terminated regions. For Si termination, the first Dy layer lies below a Si-Rh-Si triple-layer surface block which results in a reduced $4f$ photoemission (PE) signal in comparison to Dy at the surface. 
The Dy 4$f$ photoemission spectra, Fig.~\ref{PEEM_DyRh2Si2}(b), were obtained from intensity integration over the two designated regions of interest in the individual XPEEM images recorded in dependence on the kinetic energy. In Fig.~\ref{PEEM_DyRh2Si2}(b), we show Dy~4$f$ PE spectra obtained from representative regions of the bright and dark-shaded areas that are marked in Fig.~\ref{PEEM_DyRh2Si2}(a) by red- and cyan-colored squares, respectively. In the $4f$ PE spectra, we observe two different line shapes characteristic of the bright and dark areas. As we demonstrated previously, the change of the crystal electric field at the Dy surface leads to a strong reorientation of the 4$f$ moments, which are found to gain a large component within the surface plane, while they have an out-of-plane orientation in the bulk \cite{Usachov2022}. The orientation of the 4$f$ moments can be derived from a comparison of the line shapes of the Dy~4$f$ PE spectra with calculated spectra for the individual $M_J$ final states \cite{Tarasov2022}. Since the line shape of a 4$f$ PE spectrum depends also on the experimental geometry \cite{Usachov2022, Tarasov2022}, we calculated the Dy~4$f$ PE spectra for the geometry of the given XPEEM experiment.
The computational results are given in Fig.~\ref{PEEM_DyRh2Si2}(c). Comparing the experimental results in Fig.~\ref{PEEM_DyRh2Si2}(b) with the computed Dy~4$f$ spectra, we find the best agreement between the two spectra highlighted in cyan color, where the calculated one corresponds to $M_J=11/2$. This indicates that the 4$f$ moments are mostly oriented out of the (001) plane which is characteristic of the bulk. In accordance with our initial assumption, we conclude that the dark-shaded areas in the XPEEM image represent the silicide surface of the DyRh$_2$Si$_2$ crystal, where the first Dy atomic layer is hidden below the Si-Rh-Si surface block in a bulk-like environment. That the best-fitting calculated spectrum is not the one for the maximum value $M_J=15/2$ implies, however, a canting of the 4$f$ moments away from the \textit{c}~axis. In contrast, the red-colored spectrum in Fig.~\ref{PEEM_DyRh2Si2}(b) belonging to the bright area of the XPEEM image does not fit any of the calculated spectra. This indicates that the Dy surface, which is much more reactive than the silicide surface, has already undergone essential modifications due to possible contamination. Notice, that the sample has been cleaved at room temperature and was subsequently transported from the preparational to the analytical chamber. Hence, contamination of the highly reactive rare-earth-terminated surface of the crystal must be expected. 
The overall instructive results obtained from our XPEEM experiment on the (001) surface of DyRh$_2$Si$_2$ can be considered as a feasibility check for upcoming measurements. They allow us to plan further studies addressing the investigation of magnetic domains in the large family of LnT$_2$Si$_2$ materials. Based on our findings for DyRh$_2$Si$_2$, in future experiments, the focus should be laid on the chemically less active silicide surfaces. Our previously published results indicate that systems like AFM GdRh$_2$Si$_2$ may reveal surface and bulk magnetic domains below the Si-Rh-Si surface which have their own characteristic properties \cite{Guettler2016}. Our results for DyRh$_2$Si$_2$ suggest that the silicide block essentially protects the deeper-lying magnetically active lanthanide layers from experimental environments and prevents them from quick contamination and degradation in comparison to the Dy-terminated surface. This gives good hope to believe that the silicide surfaces of LnT$_2$Si$_2$ materials serve as a platform for the investigation of the physical properties and temperature scales of FM and AFM domains in such curious 4$f$-based materials by means of XMCD/XMLD PEEM measurements.


\section{Conclusions}
We determined the exchange interaction parameters of a mean-field linear-chain model by fitting those together with the CEF parameters to experimental magnetization data of DyRh$_2$Si$_2$ ($\chi(T)$ at $1\,\rm T$ and $M(H)$ at $1.8\,\rm K$). 
The simulation of the field and the  temperature-dependent susceptibility data as well as the heat capacity data shows excellent agreement with experimental data and reveals two magnetic transitions at $T_{\rm N}=55\,\rm K$ and at $T_{\rm N2}=12\,\rm K$. We studied the magnetization behavior in field and found for $H\parallel 001$ a two-step magnetization process, similar to that in HoRh$_2$Si$_2$ \cite{Shigeoka2012} and in HoIr$_2$Si$_2$ \cite{Kliemt2018}. For $H\parallel 100$, we observe the signatures of the reorientation of magnetic domains and spin-flop behavior with $H\parallel110$. The low-field $M/H$ data show that the moments are aligned closer to the $(100)$ than to the $(110)$ plane which is consistent with the result of our simulations yielding a tilting of the moments towards $[100]$ away from the $c$ direction at low T.
Finally, using XPEEM we visualized and investigated the Si- and Dy-terminated surfaces of the cleaved crystal. Although the Dy surface contaminates quickly, the silicide surface is quite robust and protects the underlying Dy layer from experimental environments. Our results suggest that the Si-terminated surface is attractive for temperature-dependent studies of magnetic domains and their properties by means of XPEEM for the huge family of strongly correlated LnT$_2$Si$_2$ materials.

\section{Acknowledgments}
We thank K.-D. Luther and T. F\"orster for technical support. We acknowledge funding by the Deutsche Forschungsgemeinschaft (DFG, German Research Foundation) via the SFB/TRR 288 (422213477, project A03) and via the SFB1143 (47310070, project C04).
This work was supported by the Saint Petersburg State University (Grant No. 94031444). D.Yu.U. acknowledges support by the Ministry of Science and Higher Education of the Russian Federation [Grant No. 075-15-2020-797 (13.1902.21.0024)]. We acknowledge the Helmholtz-Zentrum Berlin f\"ur Materialien und Energie for beamtime at the SPEEM instrument (UE49-PGM) at the BESSY II electron storage ring \cite{SPEEM}. We would like to thank Florian Kronast and Alevtina Smekhova for exceptional scientific support during the experiment.

\bibliography{bib_DyRh2Si2}

\section{Supporting Information}
\subsection{Temperature dependence of the susceptibility}
\begin{figure}[htb]
\centering
\includegraphics[width=\columnwidth]{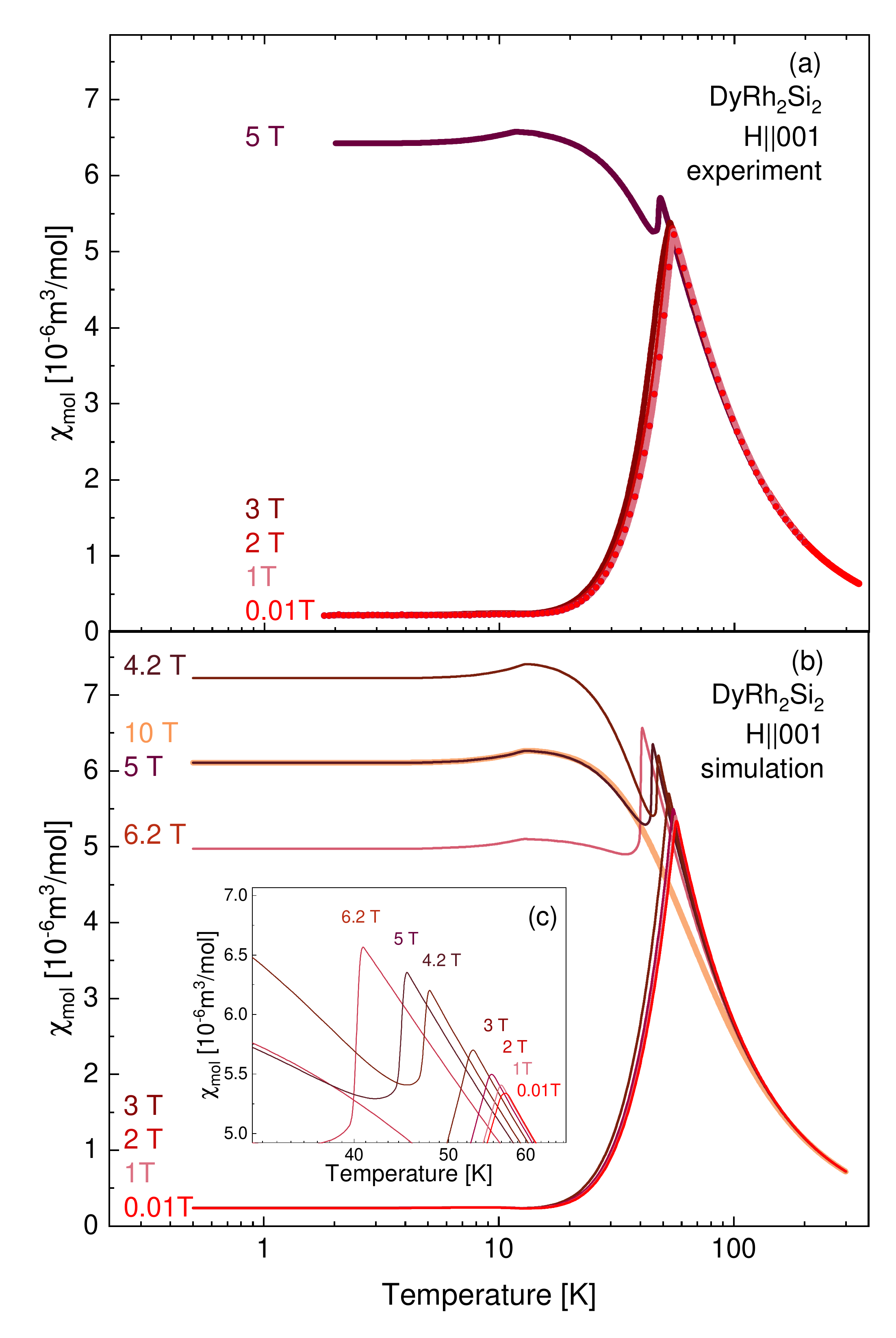}
	\caption[]{Temperature dependent susceptibility for $H\parallel 001$ with different applied fields (a) experimental (b) simulated data. In (c), an enlarged view of the simulated data close to $T_{\rm N}$ is shown.}
\label{MvT001}
\end{figure}

\begin{figure}[htb]
\centering
\includegraphics[width=\columnwidth]{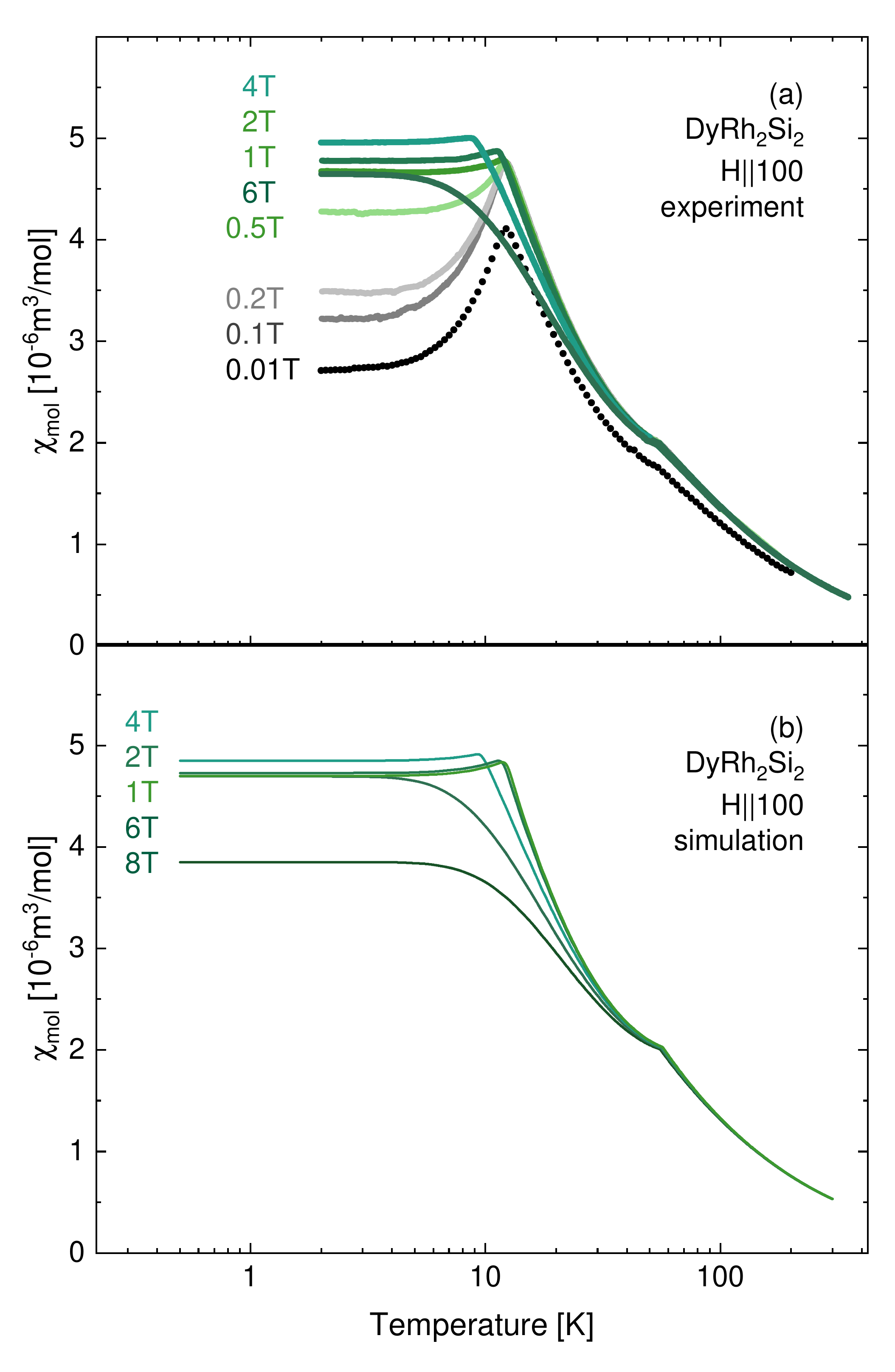}
	\caption[]{Temperature-dependent susceptibility for $H\parallel 100$ (a) experimental data and (b) simulated data.}
\label{MvT100}
\end{figure}

\begin{figure}[htb]
\centering
\includegraphics[width=\columnwidth]{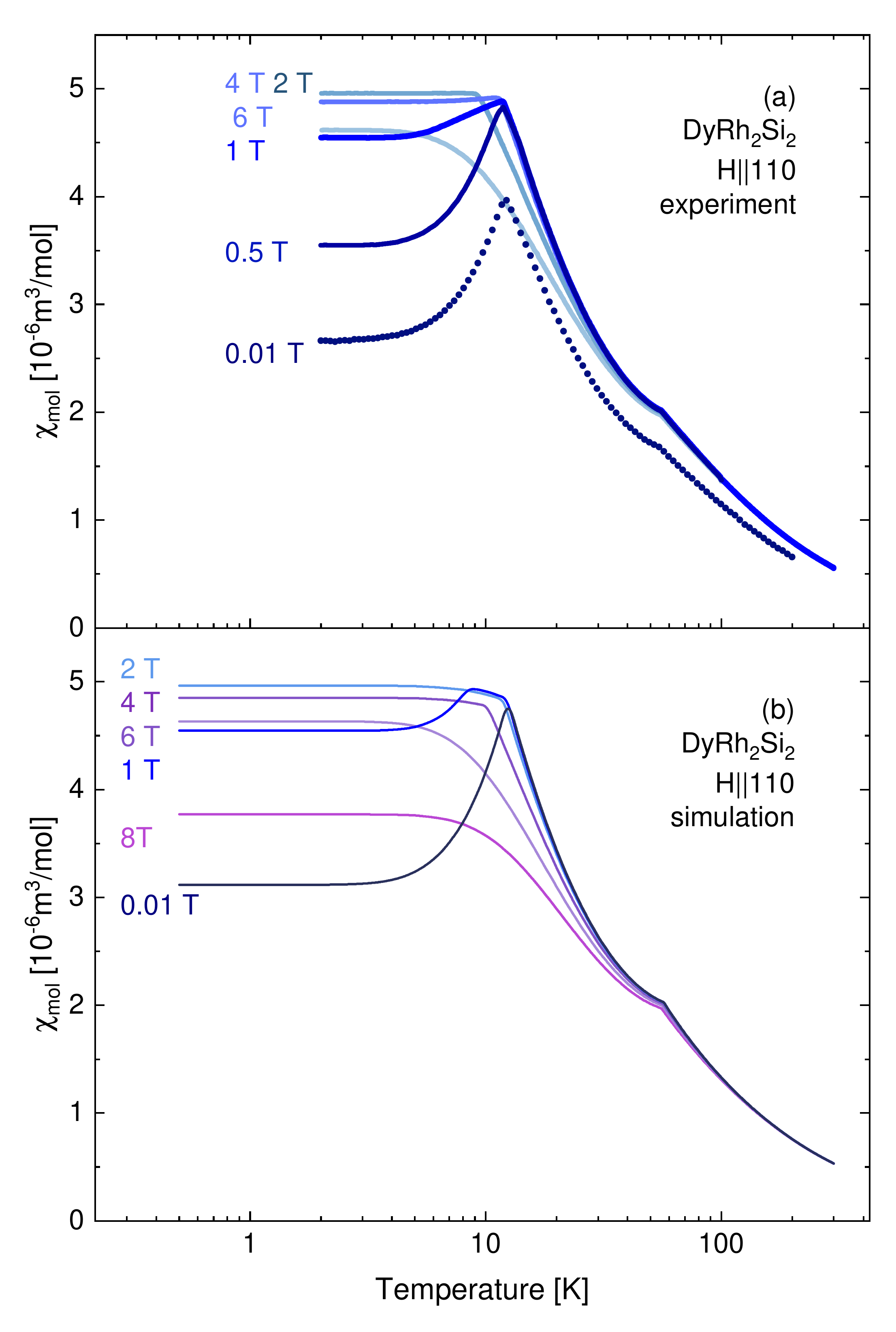}
	\caption[]{Temperature-dependent susceptibility for $H\parallel 110$ (a) experimental data and (b) simulated data.}
\label{MvT110}
\end{figure}

The experimental and simulated temperature dependence of the susceptibility for different field strengths aligned along different crystallographic orientations is shown in  
Figs.~\ref{MvT001}, \ref{MvT100} and  \ref{MvT110}.

\subsection{Field dependence of the magnetization}
\begin{figure}[htb]
\centering
\includegraphics[width=\columnwidth]{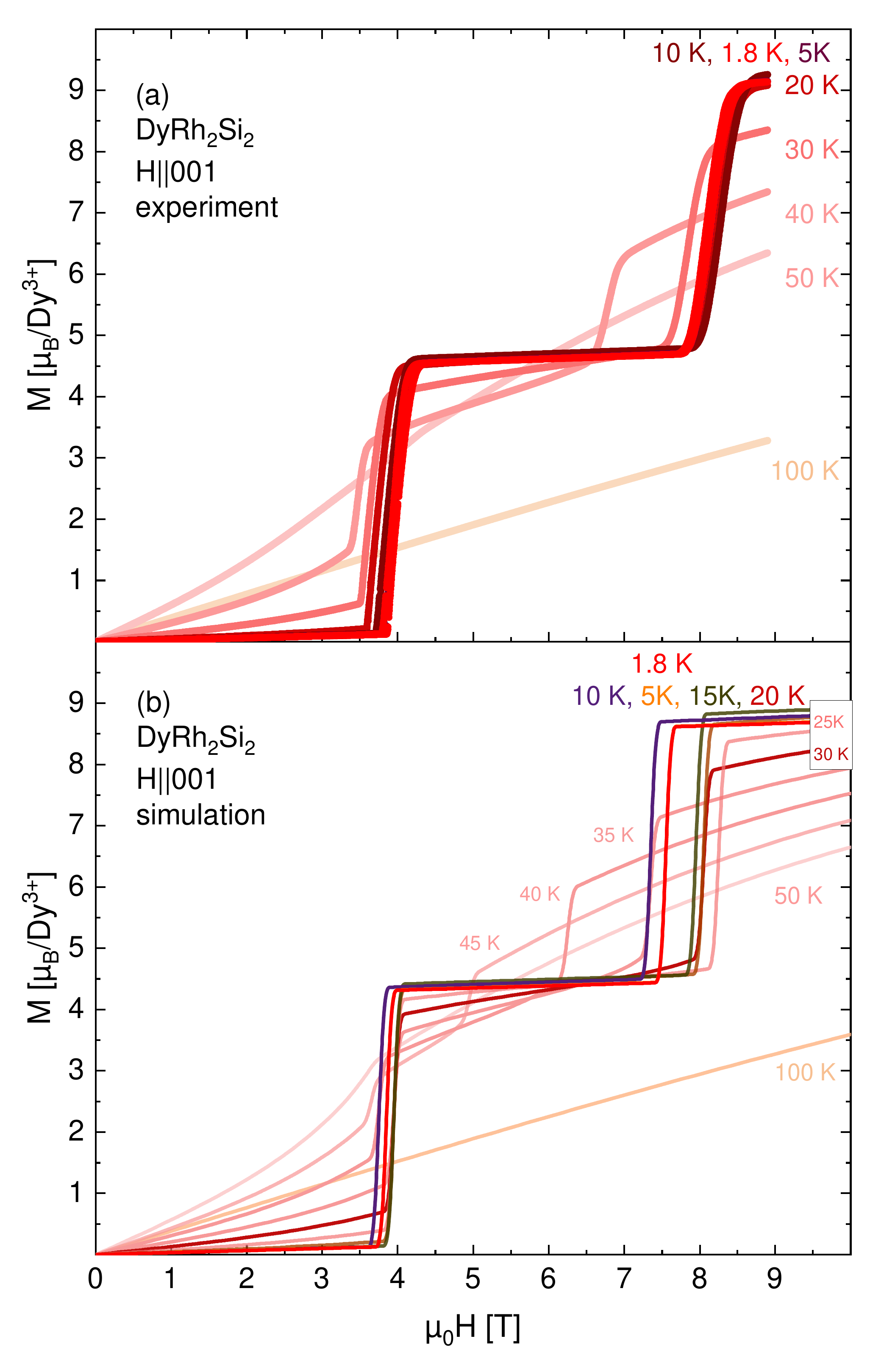}
	\caption[]{Field dependence of the moment for $H\parallel 001$ at different temperatures (a) experimental and (b) simulated data.}
\label{MvH001}
\end{figure}

\begin{figure}[htb]
\centering
\includegraphics[width=\columnwidth]{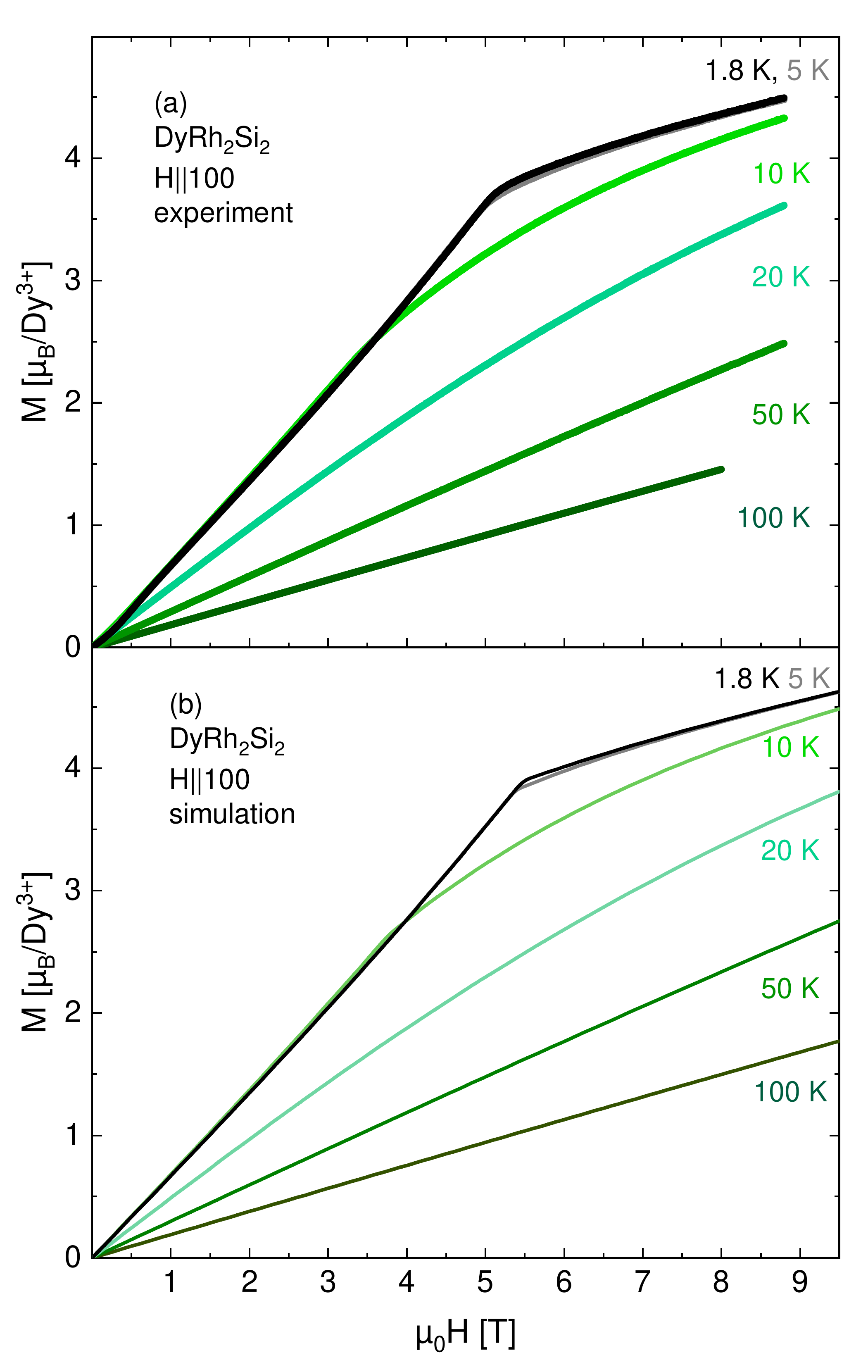}
	\caption[]{Field dependence of the moment for $H\parallel 100$ at different temperatures (a) experimental and (b) simulated data.}
\label{MvH100}
\end{figure}

\begin{figure}[htb]
\centering
\includegraphics[width=\columnwidth]{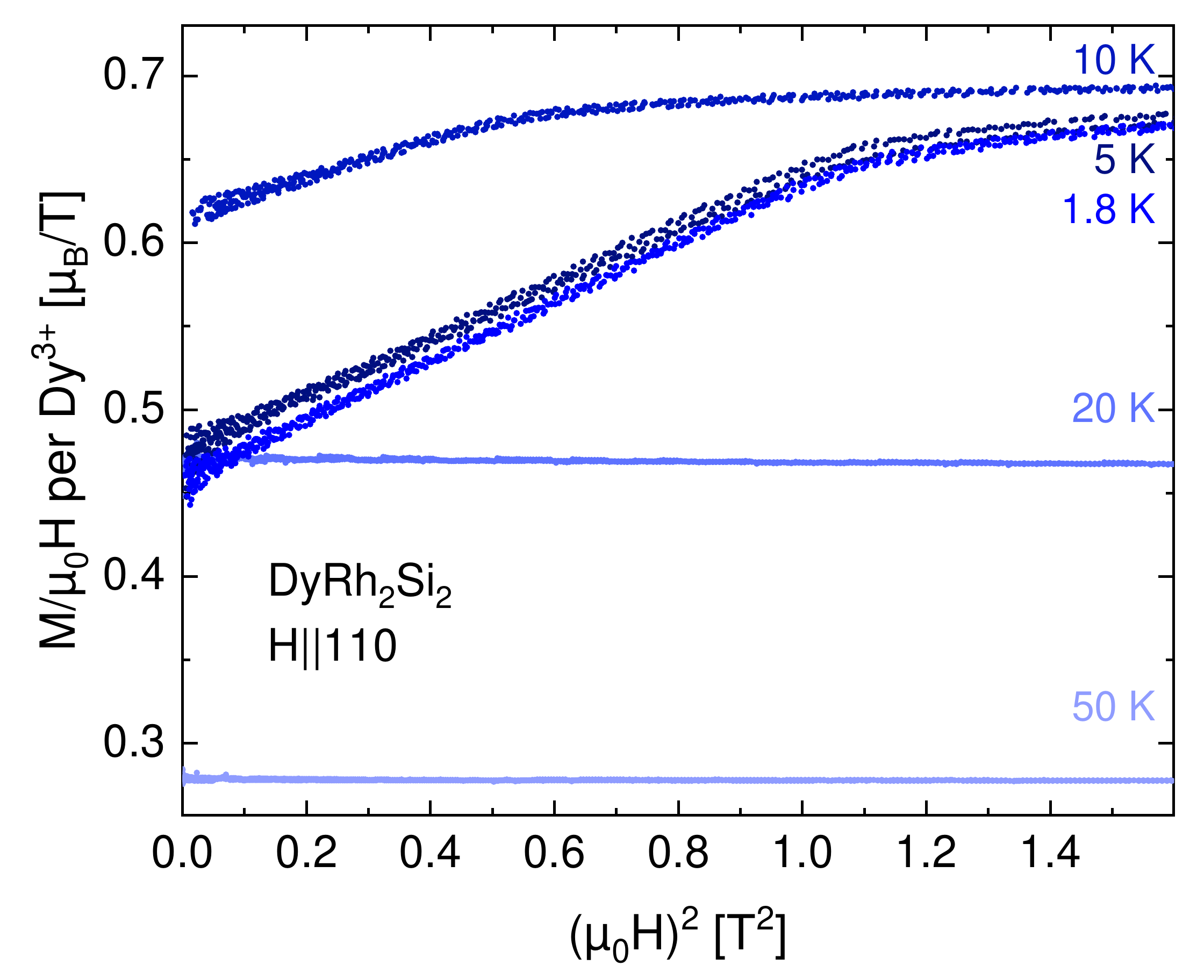}
	\caption[]{Spin-flop behavior: Field-dependence of the magnetic moment for $H\parallel 110$ for different temperatures.}
\label{MvH110_Spinflop}
\end{figure}

\begin{figure}[htb]
\centering
\includegraphics[width=\columnwidth]{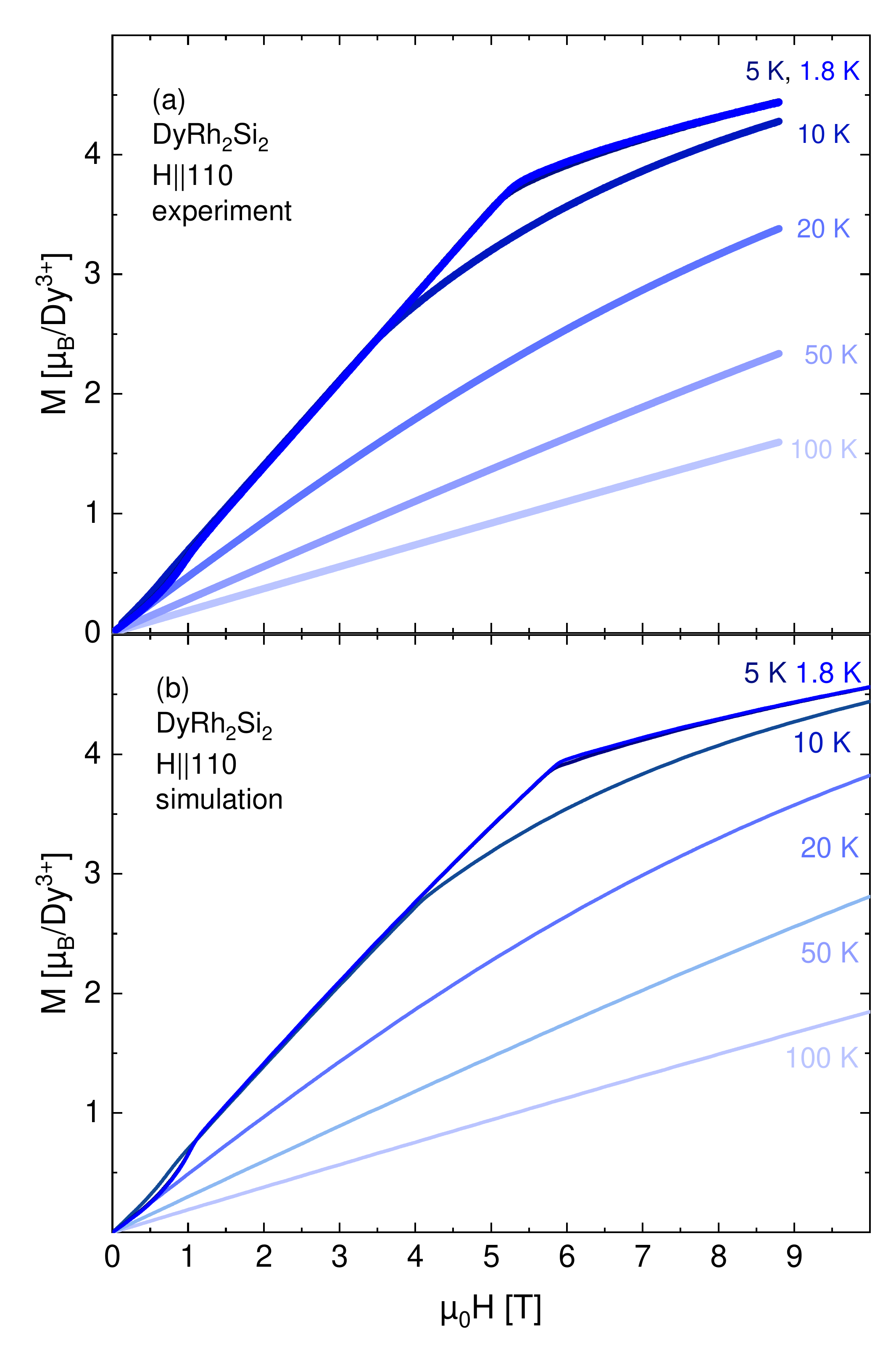}
	\caption[]{Field dependence of the moment for $H\parallel 110$ at different temperatures (a) experimental and (b) simulated data.}
\label{MvH110_all}
\end{figure}

The experimental and simulated field  dependence of the magnetic moment for different temperatures is shown in 
Figs.~\ref{MvH001}, \ref{MvH100}, \ref{MvH110_Spinflop} and  \ref{MvH110_all}.


\end{document}